\documentclass[5p,times]{elsarticle}
\usepackage[T1]{fontenc}
\usepackage{inputenc}
\usepackage[english]{babel}
\usepackage{amssymb}
\UseRawInputEncoding

\begin{document}

\begin{frontmatter}

\title{Influence of drug/lipid interaction on the entrapment efficiency  of isoniazid in liposomes for antitubercular therapy: a multi-faced investigation.}

\author{Francesca Sciolla}
\address{CNR-ISC Sede Sapienza, Piazzale A. Moro 2, I-00185 - Rome (Italy)}
\author{Domenico Truzzolillo \corref{cor1} \fnref{fn1}}
\fntext[fn1]{Laboratoire Charles Coulomb - UMR 5221, Universit\`{e} de Montpellier et CNRS, Place E. Bataillon, Campus Triolet, Batiment 11, cc 0026 34095 Montpellier Cedex 05, (France) \\ domenico.truzzolillo@umontpellier.fr}

\author{Edouard Chauveau}

\address{Laboratoire Charles Coulomb (L2C), University of Montpellier, CNRS, Montpellier, (France)}
\author{Silvia Trabalzini}
\address{Dipartimento di Chimica e Tecnologie farmaceutiche, Universit\`{a} di Roma, Piazzale A. Moro 5, I-00185 - Rome (Italy)}

\author{Luisa Di Marzio}
\address{Dipartimento di Farmacia, Universit\`{a} G.d'Annunzio, Via dei Vestini, 66100 - Chieti, (Italy)}
\author{Maria Carafa, Carlotta Marianecci}
\address{Dipartimento di Chimica e Tecnologie farmaceutiche La Sapienza Universit\`{a} di Roma, Piazzale A. Moro 2, I-00185 - Rome (Italy)}

\author{Angelo Sarra}
\author{Federico Bordi}
\author{Simona Sennato \corref{cor1} \fnref{fn2}}
\address{CNR-ISC Sede Sapienza and Dipartimento di Fisica, La Sapienza Universit\`{a} di Roma, Piazzale A. Moro 2, I-00185 - Rome (Italy)}
\fntext[fn2]{CNR ISC Sede Sapienza, Dipartimento di Fisica, La Sapienza Universit\`{a} di Roma, Piazzale A. Moro 2, I-00185 - Rome (Italy), +39 06 49913503, \\ simona.sennato@roma1.infn.it }
\cortext[cor1]{Corresponding author}

\begin{abstract}

Hypothesis.

Isoniazid is one of the primary drugs used in tuberculosis  treatment. Isoniazid encapsulation in liposomal vesicles can improve drug therapeutic index and minimize toxic and side effects. In this work, we consider mixtures of hydrogenated soy phosphatidylcholine/phosphatidylglycerol (HSPC/DPPG) to get novel biocompatible liposomes for isoniazid pulmonary delivery. Our goal is to understand if the entrapped drug affects bilayer structure.\\

Experiments.

HSPC-DPPG unilamellar liposomes are prepared and characterized by dynamic light scattering, $\zeta$-potential, fluorescence anisotropy and Transmission Electron Microscopy.  Isoniazid encapsulation is determined by UV and Laser Transmission Spectroscopy. Calorimetry, light scattering and Surface Pressure measurements are used to get insight on adsorption and thermodynamic properties of lipid bilayers in the presence of the drug.\\

Findings.

We find that INH-lipid interaction can increase the entrapment capability of the carrier due to isoniazid adsorption. The preferential INH-HSPC dipole-dipole interaction promotes modification of lipid packing and ordering and favors the condensation of a HSPC-richer phase in molar excess of DPPG. Our findings highlight the importance of fundamental investigations of drug-lipid interactions for the optimal design of liposomal nanocarriers.

\end{abstract}

\begin{keyword}
unilamellar liposomes, isoniazid, drug-lipid interaction, laser transmission spectroscopy, calorimetry, scattering techniques
\end{keyword}
\end{frontmatter}

\newpage

\section{Introduction}
Tuberculosis (TB) is caused by Mycobacterium tuberculosis (MTB), a bacterium that most often affects the lungs. The World Health Organization estimates that about one-quarter of the world's population has active or  latent TB, and that a total of 1.5 million people died from TB in 2019 \cite{world2019global}.
The current TB-treatment is usually associated with serious adverse effects, resulting in poor compliance, which is one of the main reasons
for the appearance of multidrug resistant strains and treatment's failure \cite{xu2018nanomaterials}.
Actually,  encapsulation of anti-TB drug in nanocarriers might be the modern answer for the development of innovative anti-TB strategies.  Nanocarriers can improve the efficacy of the current TB treatments since they can be functionalized to bind MTB-infected phagocites via biological ligands, and used for inhalation administration, or to enhance drug loading and pharmacokinetics, increasing significantly intracellular drug concentration.
Since earlier studies proving a ma\-cro\-pha\-ge-specific delivery of anti-TB drugs \cite{deol1997therapeutic,quenelle1999efficacy}, liposomal vesicles still remain the most widely studied carrier system for anti-TB drugs. Moreover, the possibility to nebulize of the liposomal dispersion directly into the lungs offers a powerful route to overcome the several limitations of oral and intravenous administration of anti-TB drugs \cite{marianecci2011pulmonary}.

Isoniazid (INH, pyridine-4-carbohydrazide) is one of the primary drugs used in the TB treatment  and is also well-known for its value in preventive therapy \cite{preziosi2007isoniazid}. %
Being a small hydrophilic molecule, INH is generally entrapped in liposomes by using the
film hydration method \cite{bangham1965diffusion} which leads to drug loading and retention in liposomes due to the very small drug partition coefficient \cite{rodrigues2001spectrophotometric,becker2007biowaiver}.
In general, the amount of a drug that can be entrapped in a liposome is difficult to predict, since it may depend on preparation method,  physico-chemical properties of the carrier (such as lipid composition, geometry and size) and on ionic force and pH of dispersing medium.  Actually, for any solvophilic drug, including hydrophilic ones, due to the small volume ratio between the internal volume of liposomes and that of the external medium, only a small amount of the drug molecules are encapsulated within the vesicles.

Several attempts have been described to entrap INH in liposomes with high efficiency, and since the first investigations
lipid composition has been recognized to be an important factor regulating drug loading, as well as the accumulation of liposomes in the lungs \cite{abra1984liposome}.
It has been shown that administration of sub-therapeutic INH doses entrapped in stealth liposomes composed by a mixture of phosphatidylcoline-pegylated di\-stearoyl\-phosphatidylethanolamine-cholesterol (PC-DSPE-\-PEG-\-Chol) with the anionic dicetilphosphate (DCP), is more effective and sanatory than higher concentrations of the free drug \cite{deol1997therapeutic, pandey2004liposome}.
Later, several other liposomal formulations based on DPPC, DSPC, EggPC, crude soy lecithin \cite{justo2003incorporation, gursoy2004co,chimote2010vitro,nkanga2017preparation,nkanga2017preparation}, di\-oleoyl\-pho\-spha\-ti\-dyl\-etha\-no\-la\-mine (DOPE) and DSPE-PEG \cite{kosa2016investigation} have been used for the efficient loading of INH. It's worth remarking that all these investigations considered both multilamellar and unilamellar liposomes and, being these two structures very different, a rigorous comparison of the results is difficult, especially for what concerns the entrapment efficiency.

Still, some very general principles guiding the optimization of the encapsulation process could be established. Chimote and Banerjee \cite{chimote2010vitro}, were the first to suggest that the observed high entrapment of INH ($\sim37\%$) could be attributed to the multilamellar nature of liposomes. Because of the increasing  encapsulation capability of an hydrophilic drug with increasing volume of the aqueous compartment of multilamellar vesicle, this system has been largely investigated.  A further boost to the use of multilamellar vesicles has been given by the possibility to co-encapsulate two anti-TB drugs as rifampicin and INH. The lipophilic rifampicin can be entrapped in the bilayers enclosing adjacent aqueous compartments, where INH is dispersed \cite{gursoy2004co,truzzi2019drugs}. Moreover, since the penetration depth of liposomes administered to the lungs through inhalation depends on the size of the particles in the aerosol, and particles with diameters ranging from 0.1 to 2 $\mu m$ can be effectively transported to the alveoli \cite{marianecci2011pulmonary,chimote2010vitro},  multilamellar vesicles are still under consideration.

However, despite all the aforementioned advantages offered by multilamellar structures, these are far from being ideal carriers from a biotechnological point of view, since their size and the number of their compartments cannot be controlled at will, raising precise regulatory issues \cite{Bremer2018regulatory}. For this reason, the development of optimal unilamellar liposomal vectors, able to entrap hydrophilic drugs, is highly desirable and still represents an important goal.

Interestingly, since the earliest studies, it was argued that the presence of a charged lipid could have a significant impact on the entrapment efficiency. Wasserman and coworkers \cite{wasserman1986simple} showed that the addition of a low content of the anionic Cardiolipin in a PC:Chol formulation yields a more efficient INH liposomal loading, possibly  due to a minimum of the bilayer permeability at an optimal PC:Cardiolipin stoichiometric ratio \cite{sennato2005evidence,lupi2008infrared}. Conversely, increasing cardiolipin molar fraction decreases vesicle stability \cite{wasserman1986simple}. Since the Wasserman's study dealt with multilamellar vesicles, the efficient loading found at low molar fraction of charged lipid has been explained by arguing that negatively charged bilayers produce wider aqueous spaces between lamellae, so to increase the volume of the aqueous compartments available for INH entrapment. Also, in anionic multilamellar liposomes containing DPPG and HSPC  lipids, a small amount of the anionic DPPG is able to confer stability to the vesicles, without interfering with the encapsulation and retainment of a model drug during nebulization \cite{niven1990nebulization}.
To improve mucoadhesion and nebulization performances of liposomal nanocarriers for pulmonary administration of drugs, a strategy based on polymer coating has been  explored \cite{manca2012composition, manca2014fabrication}. Also in this case, the authors report on the role of vesicle charge for optimizing the polymeric coating by electrostatic interaction with the bilayer, the charge conferring further mechanical stability to the liposomes during nebulization.

Within this framework, we focus our investigation on charged unilamellar vesicles formed by HSPC mixed with the anionic DPPG, which have not been explored as potential INH carrier so far. This formulation offers, at least, two advantages: i)
HSPC is already employed in several approved liposomal drugs \cite{zhou2015inhaled} and ii) the addition of DPPG gives the further possibility of exploiting polymeric chitosan coatings to confer mucoadhesion properties, which is relevant for pulmonary delivery \cite{m2018chitosan}.

In spite of the many different investigations on the preparation and the characterization of liposomal systems and  \textit{in vitro} and \textit{in vivo} liposomal INH delivery, some crucial aspects concerning the physical-chemical properties of the carrier are still scarcely explored. To the best of our knowledge, an inadequate attention has been paid until now to understand the interaction of INH with the lipid bilayer of the carrier. Only few biophysical investigations report on the interaction
of INH with liposomes mimicking a biological membrane, having the purpose to understand how the drug finds its way into a real membrane \cite{rodrigues_interaction_2003,chimote2008evaluation, pinheiro_interactions_2014}.

Our work leverages on an extensive characterization of the interaction between INH and mixed HSPC-DPPG  liposomes designed to optimize a novel delivery carrier for INH in anti-TB therapy. After a preliminary characterization of liposomes by dynamic light scattering, transmission electron microscopy, UV spectroscopy and laser transmission spectroscopy, which suggested the presence of drug-lipid association, we focussed on the interaction of INH with the lipid bilayers, by taking advantages of differential scanning calorimetry, static light scattering  and surface pressure measurements on Langmuir monolayers, through which we could unambiguously unveil the effect of the drug on the thermodynamics of the mixed lipid membranes.

The paper is organized by presenting the results obtained by each technique in a separate subsection. Our results support a scenario in which the interaction of INH with charged liposomes at physiological pH is affected by the fraction of charged (anionic) component of the bilayers. We had evidence of the INH permanence in proximity of the bilayer with a possible intralayer insertion, which causes modification to lipid arrangement and phase separation at high DPPG molar fraction. This represents the key finding of our investigation and we believe that it represents an important step towards a rational design of effective anti-TB liposomal nanocarriers based on the control of lipid-drug interactions.

\section{Materials and methods}\label{MatMeth}
\subsection{Materials}
\begin{figure}[htbp]
\centering
  \includegraphics[width=\linewidth]{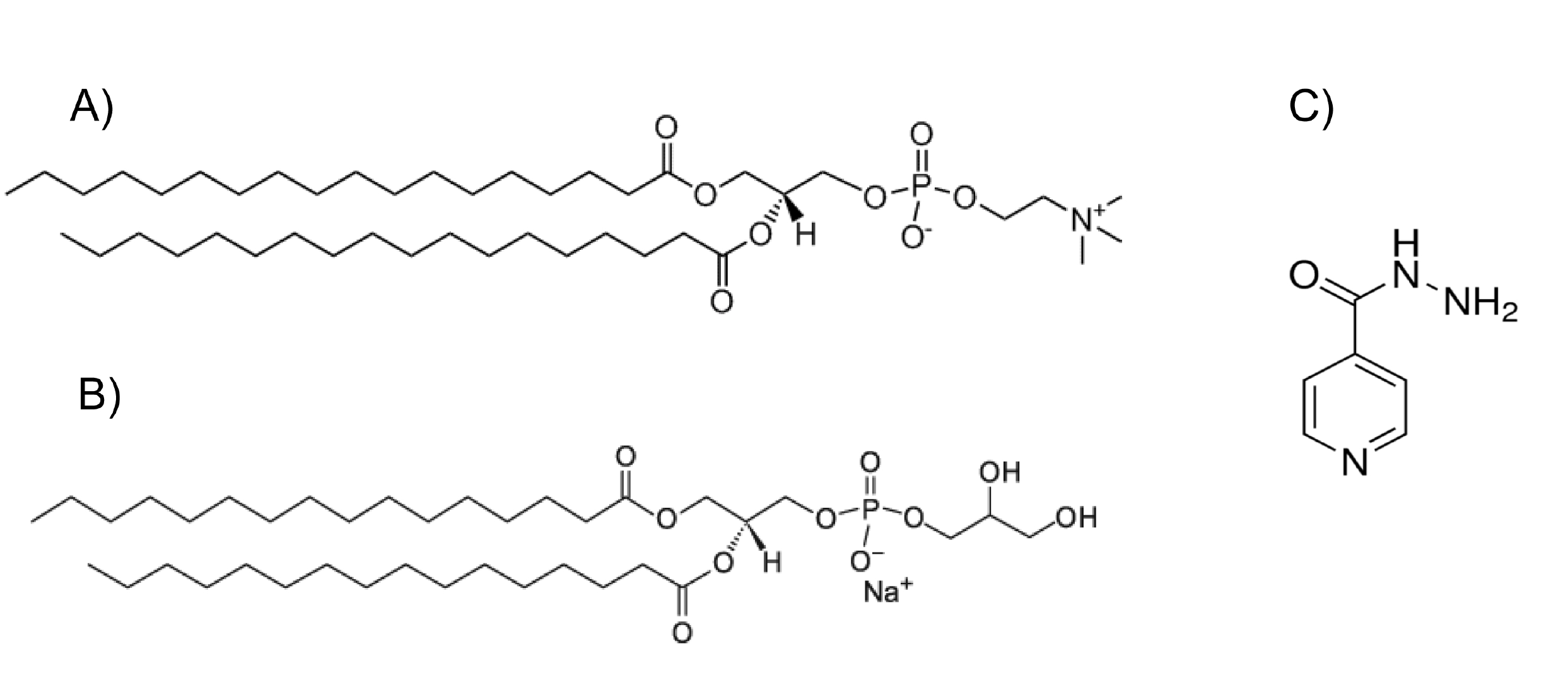}\\
  \caption{Chemical structure of  HSPC (A), DPPG-Na (B) and INH (C)}
  \label{lipids}
\end{figure}

The zwitterionic hydrogenated phosphatidylcholine from soybean (HSPC)  with molecular weight $M_w$=790 g/mol (Fig. \ref{lipids}-A) and the anionic 1,2-dipalmitoyl-sn-glycero-3-\\phosphorylglycerol sodium salt (DPPG) with molecular weight  $M_w$=745 g/mol (Fig. \ref{lipids}-B) were a kind gift from LIPOID. The typical fatty acid composition (expressed in \% of total fatty acids) of HSPC is:  palmitic acid: (5.0\% - 20.0 \%) and stearic acid (80.0 \% - 95.0 \%).

Hepes salt [N-(2- hydroxyethyl), piperazine-N-(2-e\-tha\-ne\-sul\-pho\-nic a\-cid)]  and isoniazid (pyridine-4-carbohydrazid, nominal purity 99\%, hereinafter INH), were purchased by Sigma Aldrich. Se\-pha\-dex G-50$^{TM}$ has been purchased from GE - Healthcare. The chemical structure of INH is shown in Fig. \ref{lipids}-C. From a chemical point of view, INH has three pK$_a$ values, 1.8 for the basic pyridine nitrogen, 3.5 for the hydrazine nitrogen and 10.8 for the hydrazine group and it is neutral at physiological pH \cite{pinheiro_interactions_2014}.
The drug was dissolved in  0.01 M Hepes buffer at pH values of 7.4,  prepared with Milli-Q grade water with pH adjusted with NaOH addition.

\subsection{Preparation of liposomes}
Lipids were dissolved in a known volume of chlo\-ro\-form\-/me\-tha\-nol\-/wa\-ter (2/1/0.15 v/v/v) at varying DPPG molar fraction $X_{PG}=n_{HS}/(n_{HS}+n_{PG})$, where $n_{HS}$ and $n_{PG}$ are the number of HSPC and DPPG moles, respectively. A three-hour rotoevaporation of the solvent under vacuum and above the melting temperature $T_m$ of both lipids resulted in the formation of a dried lipid film. By rehydration of the lipid film in Hepes 0.01 M solution and pH=7.4, through a uniform rotation at $T > T_m$ for one hour, a dispersion of multilamellar liposomes was obtained. For calorimetry measurements, multilamellar liposomes were prepared at a lipid concentration of 25 mg/mL.

In order to obtain unilamellar vesicles, the hydrated lipid suspension was subsequently homogenized by 5 cycles of freeze-thaw and extruded 10 times under nitrogen pressure through a 100 nm polycarbonate membrane (Whatman Nucleopore) in a 2.5 mL extruder (Lipex Biomembranes, Vancouver, Canada) at 60 ${^\circ}$C, well above the main transition temperature of lipids. Unilamellar liposomes for calorimetry and light scattering experiments have been prepared at 10 mg/mL and 0.2 mg/mL, respectively.

To entrap INH, the dried lipid film has been hydrated using Hepes buffer containing the drug dissolved at the target concentration. As for empty liposomes, 5 freeze-thaw cycles have been applied, since this procedure is also able to facilitate encapsulation of hydrophilic drugs \cite{xu2012predicting}. The non-entrapped INH was separated from the liposomes on a Sephadex G-50 gel column hydrated in Hepes buffer after 24 hours of swel\-ling. The amount of liposomal solution to be purified was assessed in dependence of the concentration of the entrapped drug and was set to 100-200 $\mu$L of liposome solution for 2.5 mL of gel.

\subsection{Dynamic light scattering and electrophoretic mobility measurements}
The size and the size-distribution of liposome formulations entrapping INH were analyzed via dynamic light scattering (DLS) measurements performed with a  NanoZetaSizer apparatus e\-quip\-ped with a 5 mW HeNe laser (Malvern Instrument, UK). This instrument employs a backscatter detection, i.e. the scattered light is collected at an angle of 173$^o$ . The main advantage of this detection geometry, when compared to the more conventional 90$^{\circ}$, is that it is less sensitive to multiple scattering effects \cite{dhadwal1991fiber}. Decay times $\tau$ were used to determine the diffusion coefficients $D_0=1/(\tau q^2)$ of the particles, which in turn can be converted in apparent hydrodynamic radii $R_h$, using the Stokes-Einstein relation $R_h = k_BT/6\pi\eta D_0$. In the above relations $q = 4\pi n \lambda^{-1} \sin(\theta/2)$ is the scattering vector, $n$ the solvent refractive index, $\lambda$ is the light wavelength, $\theta$ the scattering angle,  $k_B$ the Boltzmann constant, $T$ the absolute temperature and $\eta$  is the solvent viscosity. The measured autocorrelation functions were analyzed using the cumulant methods to get the mean hydrodynamic size and the polidispersity index (PDI) by the first and second moment of the cumulant expansion, respectively \cite{koppel1972analysis}. Results are expressed as the average of three different measurements, each measurement was averaged over at least 20 runs.

By using the same apparatus, the electrophoretic mobility has been measured to determine their $\zeta$-potentials. The mobility $\mu$ of the liposomes is  converted into a  $\zeta$-potential using the Smoluchowski equation  $\zeta = \mu \eta/ \epsilon$, where $\epsilon$ and $\eta$ are respectively the zero-frequency absolute dielectric permittivity and the viscosity of the suspending medium.

\subsection{Laser Transmission Spectroscopy}\label{sec:LTS}
The size and the absolute number concentration of the liposomal suspension  were determined using an innovative and customized apparatus implementing the laser transmission spectroscopy (LTS) technique \cite{li2010high,de2019balanced}. Since it is relatively new and probably unfamiliar to the readership, we will  give here a very  brief account of this technique.  By  measuring the light  transmittance through the vesicle suspension as a function of the wavelength,  the particle density distribution $n(R)$ as a function of their size $R$ can be obtained through the Beer-Lambert law
once the Mie scattering cross section of the vesicles, represented as shelled spheres, is known \cite{li2010high}. For this purpose we used a pulsed laser tunable in the wavelength interval from 210 to 2600 nm.
Transmission data are analyzed and inverted by using a mean square root-based algorithm, giving the particle size
distribution in terms of their absolute concentration. The integral of the density distribution provides
the total number of liposomes per milliliter of solution $N_{LTS}$ \cite{de2019balanced,de2020blueberry}.  The volume fraction $\Phi_{in}$ of the liposomal dispersion available for encapsulation can be hence calculated as  $\Phi_{in}= N_{LTS} \cdot 4/3 \pi (R-d)^3$, where $d$ is the bilayer thickness.

\subsection{Transmission electron microscopy}
Transmission electron microscopy (TEM) measurements  we\-re carried out by using a FEI TECNAI 12 G2 Twin
(Thermo Fisher Scientific - FEI Company, Hillsboro, OR, USA), operating at 120 kV and equipped with an electron energy loss filter and a slow-scan charge-coupled device camera (794 IF, Gatan Inc, Pleasanton, CA, USA). 20 $\mu$l of the sample have been deposited on 300-mesh copper grid covered by thin amorphous carbon film. After 2 minutes the excess liquid was removed by touching the grid to filter paper and 10 $\mu$l of 2 $\%$ aqueous phosphotungstic acid (PTA)  (pH-adjusted to 7.3 using 1 N NaOH) has been added to stain the sample. 

\subsection{Lipid bilayer characterization by fluorescence anisotropy}
Fluidity of liposomal bilayers was evaluated  by inspecting fluorescence anisotropy of  diphenylhexatriene (DPH) probe dissolved in the hydrophobic region of the lipid bilayer. 250 $\mu$L of DPH (2 mM)  was added to lipid mixtures before vesicle preparation according to a protocol already discussed in a previous study \cite{marianecci2013polysorbate}. Afterwards, DPH-loaded vesicles were extruded as done for empty liposomes.
Fluorescence anisotropy was measured by a LS55 spectrofluorimeter (PerkinElmer, MA, USA) at $\lambda_{exc}$=400 nm and $\lambda_{em}$= 425 nm, at room temperature. The fluorescence anisotropy (A) of samples was calculated according to the following equation
\begin{equation}\label{eq:anisotropy}
    A= \frac{I_{vv}-G I_{vh}}{I_{vv}+2G}{,}
\end{equation}
where $I_{vv}$ and $I_{vh}$ are the intensities of the emitted fluorescence (arbitrary units) parallel and perpendicular to the direction of the vertically polarized excitation light, respectively. $G = I_{vh}/I_{hh}$ is a correction factor, which is determined experimentally before each measurement as the ratio between the vertically and the horizontally polarized emission components, for a given horizontally polarized incident beam. The fluorescence a\-ni\-so\-tro\-py values A are inversely proportional to the membrane fluidity, high values of A correspond to a high structural order and/or a large viscosity of the membrane \cite{shinitzky1978fluidity}.

\subsection{Entrapment efficiency }\label{drugloading}
Quantification of INH loaded in liposomal formulations was carried out by a UV-VIS Jasco
spectrophotometer with 1 mm quartz cuvettes, at 20.00 $^\circ$C. To subtract the contribution of background scattering, spectra of empty liposomes measured in a wide concentration range (1 to 20 mg/ml) have been collected, with Hepes buffer as reference. Preliminarily, two different methods of background subtraction have been tested: i) spectra of INH-loaded liposomes measured with empty liposomes as reference, at the same lipid concentration; ii) spectra of INH-loaded liposomes measured with Hepes as reference, from which the spectra of empty liposomes, at the proper concentration and with Hepes as reference,  have been subtracted. Since the obtained background-subtracted UV spectra were e\-qual,  we have chosen method (ii) for convenience.
Drug concentration before and after purification have been determined by considering the INH absorption maximum at $\lambda\sim$ 264 nm present in the background-subtracted spectra \cite{barsoum2008spectrophotometric}, and comparing the intensity with a calibration curve obtained in Hepes buffer.

Entrapment efficiency (E.E.) has been calculated according to the following equation
\begin{equation}\label{eq:EE}
    E. E.(\%) =  \frac{C^{f}_{INH}/C^f_{lipid}}{C^{0}_{INH}/C^0_{lipid}}\cdot100 {,}
\end{equation}
where the concentrations $C^0_{INH}$, $C^{0}_{lipid}$ and $C^f_{INH}$,  $C^{f}_{lipid}$  are referred to the  molar  concentration of INH or lipid, before and after purification, respectively.

\subsection{Differential scanning calorimetry} \label{sub:DSC}
Differential scanning calorimetry (DSC) experiments were performed with a TA Q2000 DSC calorimeter. The measurements were carried out under nitrogen flow. A modulated temperature protocol with an amplitude of 0.3 $^{\circ}$C over a period of 60 s was applied within heating/cooling ramps in the temperature range 10 $^{\circ}$C $\leq T\leq$ 70 $^{\circ}$C at 2 $^{\circ}$C/min. For each sample we performed 3 heating/cooling cycles to erase any past thermal history of the sample and achieve stationary thermograms. Excess molar calorimetric enthalpy was calculated after baseline adjustment and normalization to the lipid concentration, by integrating the peak areas. The actual transition temperatures ($T_m$) were determined at the peak maxima of the heat capacity curves.

To investigate the interaction of INH with the lipid layer, liposomes with different concentrations of INH have been prepared by mixing a fixed volume of drug (10 $\mu$l) to 500 $\mu$l of pure unilamellar liposome suspension,  for each molar fraction $X_{PG}$. INH concentration was varied to obtain a final INH/lipid molar ratio $\rho$ ranging from 0.5 to 25.

\subsection{Static light scattering}
The static light scattering  (SLS) experiments have been performed using an Amtec-goniometer equipped with a green laser with wavelength $\lambda$ = 532.5 nm. All measurements were performed at the same incident light intensity (laser power set at 70 mW).  The temperature $T$ of the cell was set by means of a temperature-regulated bath with an accuracy of 0.1 $^{\circ}C$.
We measured the gyration radius of the liposomes by collecting the scattered intensity $I(q)$ scattered by dilute samples (0.2 mg/ml) at 60 scattering angles $\theta$ between $24^\circ$ and $150^\circ$.  From the time averaged scattering intensity $I(q)$ the radius of gyration $R_g$ has been determined by using the Guinier approximation $I(q)=I(0)\exp[-(qR_g)^2/3]$ \cite{berne_dynamic_1990}.

By fitting  the $R_g(T)$ curves, or the  time-averaged intensity at $\theta$=$90^\circ$,  with a Boltzmann-type equation, it is possible to determine the (optical) melting transitions $T_c^{opt}$  \cite{michel_determination_2006}
\begin{equation}\label{Boltzmann2}
    R_g(T)=(R_s-R_{m})\cdot\left( 1+e^{\frac{T-T_c^{opt}}{\Delta T}} \right){,}
\end{equation}
where $R_s$ and $R_m$ represent the gyration radii in the solid and melted state, respectively, and $\Delta T$ is the transition width.

A proper drug amount to get a INH/lipid molar ratio $\rho=5$ was added to each pure unilamellar liposome suspension, to investigate the effect of INH on the lipid bilayer.

\subsection{Monolayer studies}
Surface pressure measurements have been performed by a Minitrough (KSV Instruments Ltd, Helsinki, Finland) equipped with Wilhelmy-type pressure measuring system, enclosed in a plexiglas box to reduce surface contamination. Hepes  solution (0.01M, pH=7.4) thermostatted at   $25.0\pm~0.2$~$^{\circ}$C has been used as subphase.

Lipid monolayers with different molar fractions of HSPC and DPPG were prepared at the air-water interface according to
the Langmuir technique \cite{Roberts90} as described in previous investigations \cite{bordi2006charge}. Lipids were dissolved in chloroform at 1 mg/ml and an amount of $20\div25 \mu l$ was spread by a microsyringe onto the aqueous subphase. After evaporation of the solvent, the monolayers were compressed by the two barriers moving at a constant rate of 50~mm min$^{-1}$ to record the pressure/area ($\Pi/ A$) isotherm. The reported isotherms represent the average over three different compression experiments.

For drug-lipid interaction studies, INH has been injected in the subphase under an already formed lipid monolayer. INH has been dissolved in ethanol at its maximum solubility concentration ($\sim$ 67 mM) and its concentration in the subphase has been varied by changing the injected volume.
Ethanol is less dense than water and rather highly  volatile and favors INH spreading at the air-water interface.
Control experiments were further performed by injecting pure ethanol in the subphase under the monolayer to quantify the extent of surface pressure increase due to pure ethanol in function of the injected volume. All the isotherms were recorded after waiting for the stabilization of the surface pressure.

The miscibility of HSPC and DPPG can be analyzed by calculating the excess of free energy of mixing $\Delta G$  upon integration of the surface pressure-area ($\Pi$-A) isotherms, from zero to a certain value of the surface pressure, according to the expression
\begin{equation}\label{eq:DG}
    \Delta G = \int_0^{\pi} (A_{12}-X_1A_1-X_2A_2) d \pi {,}
\end{equation}
where $A_i$ and $X_i$ are the area per molecule and the molar fraction of component \textit{i}, respectively, and $A_{12}$ is the area per molecule
in the mixture. In the absence of any interaction between the components, $\Delta G= 0$. Deviations from an ideal behavior results
in $\Delta G < 0$ (attractive interactions) or in $\Delta G> 0$ (repulsive interactions), providing information on whether the interaction
is energetically favored or not \cite{sennato2005evidence}.

\section{Results}
\subsection{Basic characterization of HSPC-DPPG liposomes}
\begin{figure}[htpb]
\centering
  \includegraphics[width=0.6\linewidth]{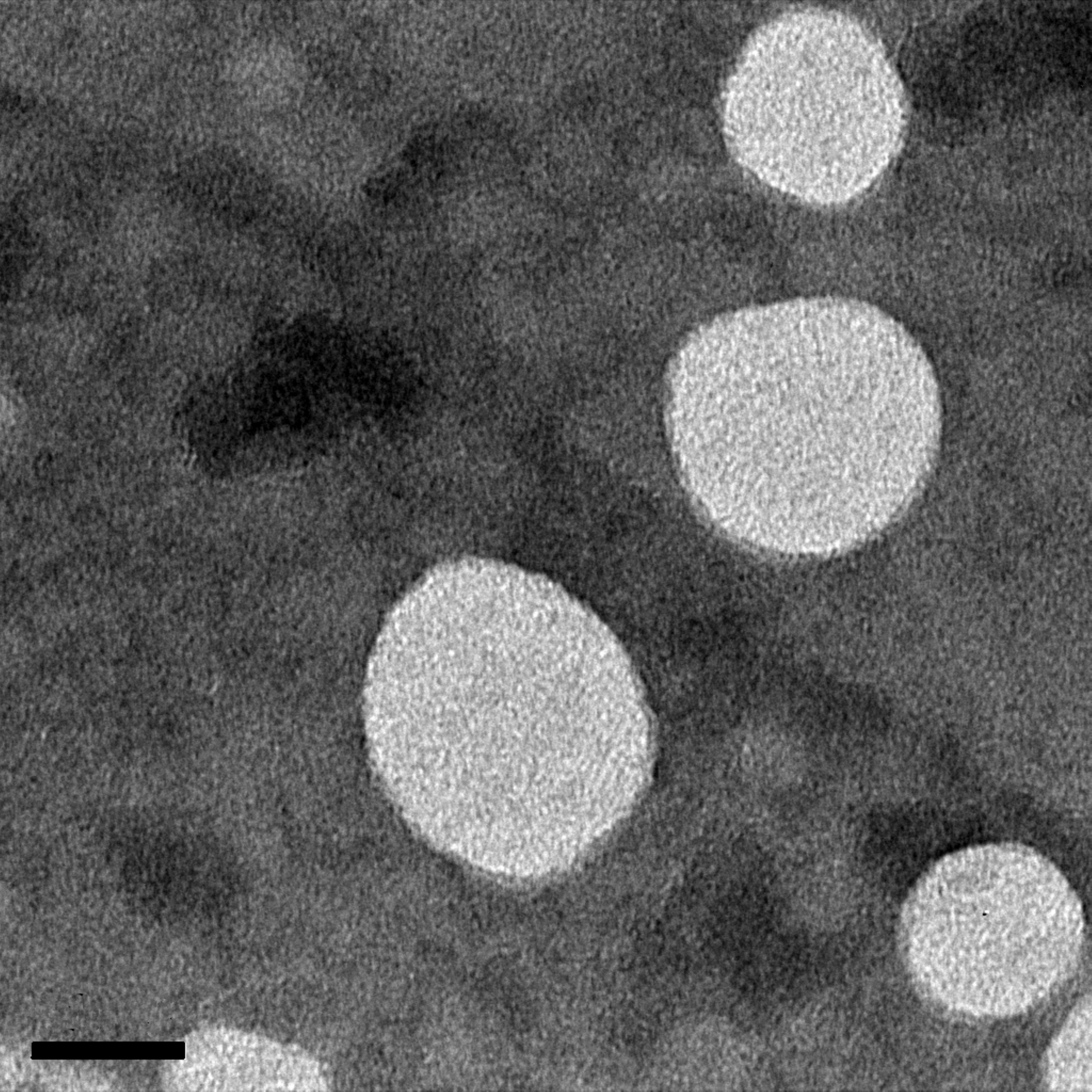}\\
  \caption{TEM microscopy image obtained by PTA staining of unilamellar liposomes prepared at $X_{PG}$= 0.66. The bar length corresponds to 50 nm.}\label{fig:TEM}
\end{figure}

Size, polydispersity, $\zeta$-potential and fluorescence anisotropy of empty unilamellar liposomes have been measured for three different $X_{PG}$ molar fractions: 0.33, 0.5 and 0.66 (Table \ref{tab:sz-emptylipo}). We obtain liposomes with size around 100 nm for all the different compositions and a low polydispersity index (PDI$\leq$0.1).
TEM microscopy confirms the presence of a homogeneous population of unilamellar vesicles, as shown in Fig. \ref{fig:TEM} for  liposomes prepared with $X_{PG}$= 0.66.

\begin{table}[htpb]
\begin{center}
\footnotesize
\caption{Characterization of empty HSPC-DPPG unilamellar liposomes at $T$=25 $^{\circ}$C and E.E. values for liposomes prepared at INH/lipid molar ratio=10 (200 mM INH). Results represent the mean values over three repeated measurements. No significant variation has been observed on size and  $\zeta$-pot values of INH-loaded liposomes (data not shown). }
\label{tab:sz-emptylipo}
\begin{tabular}{cccccc}
$X_{PG}$ & $R_h (nm)$     & PDI             & $\zeta$-pot (mV)&    Anisotropy     & E.E. (\%)       \\
\hline
0.33     & 57 $\pm$ 3     & 0.08 $\pm$ 0.01 & -35 $\pm$ 4     & 0.37 $\pm$ 0.04&   $2.4  \pm  0.2$   \\
0.50     & 55 $\pm$ 2     & 0.07 $\pm$ 0.01 & -53 $\pm$ 6     & 0.35 $\pm$ 0.05&   $1.7  \pm   0.2$    \\
0.66     & 52 $\pm$ 2     & 0.11 $\pm$ 0.02 & -42 $\pm$ 4     & 0.37 $\pm$ 0.03&    $1.1  \pm   0.1$  \\
\end{tabular}
\end{center}
\end{table}
The values of $\zeta$-potential are negative for all the three formulations, due to the presence of the anionic lipid DPPG.  $\zeta$-potential does not show a linear increase with increasing charged lipid content since it
is not connected with the stoichiometric charge but with the effective charge of the system \cite{bordi2006charge}.
As it has been shown for other charged liposomal systems, while the stoichiometric  charge increases linearly with the charged lipid content, the effective charge of liposomes displays a saturation due to the counterion condensation which minimizes the repulsion between the nearby negative charges \cite{bordi2006charge}. Moreover we expect that the $\zeta$-potential is affected by the dipolar orientation of the zwitterionic head groups \cite{szekely2011structure, zimmermann2009charging}, that are susceptible to any variation of the local composition within the membrane. However, this aspect, interesting \emph{per se}, goes well beyond the scope of our work and has not been investigated further.

The observed high values of anisotropy point out that all the formulations have a rather rigid lipid bilayer, as the pure components DPPG, DPSC and DPPC liposomes \cite{zuidam1995physical}. This is indeed expected since both lipids are in a gel state at 25 $^\circ$C.

We performed a preliminary set of experiments to determine the best operative condition for encapsulation of INH within the range $1\div 50$ of INH/lipid molar ratio (see SI, section 1).   In Table \ref{tab:sz-emptylipo}  we reported the highest values of entrapment efficiency which have been found in the presence of an intermediate excess of drug, i.e. at  INH/lipid ratio equal to 10. In general, we found low values for the $E.E.\%$  parameter. The  highest value of E.E. parameter is found at lower content of the charged lipid DPPG and  a decreasing behavior with increasing molar fraction $X_{PG}$ is observed. This could indicate a worse retention capability of the vesicles due to the looser packing in more charged bilayers. In fact, it is known that the presence of electrostatic repulsion between charged lipids may alter the bilayer properties and increase permeability to solutes \cite{okahata1984bilayer}. However, this hypothesis does not appear suitable since anisotropy values of HSPC-DPPG liposomes close to 0.36 indicate a rather rigid bilayer. Moreover, it has been shown that INH does not induce any change in fluidity in DMPC and DMPG liposomes \cite{rodrigues_interaction_2003}, which are more fluid and with a lower melting temperature compared to the lipid used in the present study.

From a purely operative point of view, the values of the E.E. parameter are useful to select the more convenient mixture to get the highest amount of entrapped INH,  given a fixed amount of lipid mass and drug used to prepare the samples. Such an optimal mixture here is reached for $X_{PG}$=0.33.
On the other hand, the only determination of E.E. does not allow unveiling how the amount of entrapped drug is influenced by chemical-physical properties of the lipid bilayer, namely, by the size and total volume of the carriers available for drug encapsulation, and by the lipid composition of the carriers.

\subsection{LTS study  of INH-loaded liposomes}
To get further insight in the encapsulation properties of HSPC-DPPG liposomes, we performed LTS experiments on INH-loaded liposomal suspensions prepared at INH/lipid molar ratio equal to 10, where the highest values of drug entrapment have been found regardless of lipid composition. The aim of this study is to determine the radius $R$ and the total number of liposomes per milliliter of solution $N_{LTS}$, and then calculate the liposome volume fraction available for drug encapsulation $\Phi_{in}$, as described in section \ref{sec:LTS}. Once the volume fraction is obtained, it is possible to define a new parameter named 'Entrapment ratio'  (hereinafter E.R.) as the ratio between the concentration of drug encapsulated in the vesicles (determined by UV) and the maximum amount of drug which can be loaded in their internal volume
\begin{equation}\label{eq:ER}
    E.R. =  \frac{C^f_{INH}}{C^0_{INH}\cdot \Phi_{in}} = \frac{E.E.\%}{100 \cdot \Phi_{in}}{.}
\end{equation}
Thank to this definition, E.R. can give indications on the entrapment efficacy of a lipid membrane with a specific lipid composition, independently of its geometrical features. We will show that the E.R. is of valuable help for our analysis.

The values of $E.R.$ are shown in Table \ref{tab:encap1} and have been calculated by assuming a bilayer thickness $d$ of 5 nm for all the HSPC:DPPG mixtures, corresponding to the average value reported for pure DSPC and DPPG bilayers \cite{xu2012predicting}.
\begin{table*}[htbp]
\begin{center}
\caption{Results of LTS study of INH-loaded liposomes, prepared at molar ratio $INH/lipid =10$  (200 mM INH). $R$ is the radius of liposomes, $N_{LTS}$ is the total number of liposomes per milliliter of solution, $INH/lip$ is the liposomal amount of INH, $\Phi_{in}$ is the liposome volume fraction,  $E.R.$ is the entrapment ratio calculated according to eq. \ref{eq:ER}.}
\label{tab:encap1}
\footnotesize
\begin{tabular}{cccccc}
$X_{PG}$ & R (nm)   & $N_{LTS} (part/ml)$       &  $INH/lip \,(mM/part)$  &     $\Phi_{in}$                      & E.R.   \\
\hline
0.33  & $45 \pm 4$ & $(2.7 \pm 0.2)\,10^{12}$ &  $(1.8 \pm 0.2)\,10^{-16}$&  $(0.73\pm 0.07)\,10^{-3}$  &  $3.3 \pm 0.3 $  \\
0.50  & $40 \pm 4$ & $(1.0 \pm 0.1)\,10^{13}$ &  $(1.4 \pm 0.2)\,10^{-16}$& $(1.8 \pm  0.2) \,10^{-3}$&  $3.8  \pm   0.4 $\\
0.66  & $37 \pm 3$ & $(6.6 \pm 0.6)\,10^{12}$ &  $(0.9 \pm 0.1)\,10^{-16}$& $(0.91 \pm 0.09)\,10^{-3}$&  $3.3  \pm    0.3$\\
\end{tabular}
\end{center}
\end{table*}

LTS  measurements show that the radius of liposomes determined by this technique has a similar behavior to the hydrodynamic radius determined by  DLS technique (see table \ref{tab:sz-emptylipo}) but it is shifted to smaller values, as expected since LTS method gives the distribution of the geometrical sizes of the suspended particles and not their equivalent 'hydrodynamic' radii (i.e. radius of a sphere with the same diffusion coefficient) \cite{de2020blueberry}.

Thanks to the determination of the particle concentration of the liposomal suspension ($N_{LTS}$), we can go a step further and calculate the 'INH liposomal amount' ($INH/lip$), i.e. the amount of INH loaded in each liposome,  by dividing the total concentration of INH determined by UV measurements 
with respect to $N_{LTS}$, assuming a uniform drug repartition in the suspension.
We note that the values of $INH/lip$ decrease with increasing the molar fraction of the charged  lipid DPPG, this could be due to the geometrical radius which decreases, too. At a single-nanocarrier level, we can conclude that liposomes with the lowest content of the charged lipid are able to retain more INH. Both lipid composition and liposome size could affect the  entrapment capability of the single nanocarrier.

This said,  to filter out the effect of the geometrical features of the liposomes, namely the liposomal volume and the number of liposomes in each suspension, we can perform a further analysis of the properties of drug encapsulation by considering the E.R. parameter. First, it is noteworthy that $E.R.$  is always higher than unity, that is the value corresponding to the absence of drug-lipid interaction and drug leakage \cite{xu2012predicting}.  Then, we note that the E.R. values are rather similar and comprised between 3 ad 4. Considering that in the pre-purified liposomal dispersion INH is added in large excess,  the  high E.R. values mean that vesicles can retain around them and/or within lipid bilayers  about three times more molecules than the amount of drug expected on the basis of the geometrical volume. More, our  findings suggest the preferential interfacial localization of INH.

In the equimolar mixture, the E.R. value is slightly higher than in the asymmetrical formulations. By its definition,  E.R. helps us to establish which formulation gives the highest drug entrapment with respect to the available volume of the suspension. In other words, the analysis of E.R. allows to compare the different liposomal suspensions as if they were composed by the same number of identical vesicles to give evidence to the influence of bilayer properties in drug entrapment. E.R. results give evidence that mixed liposomal vesicles at equimolar HSPC-DPPG  ratio are the most effective for INH entrapment.

To sum sup, our findings indicate the presence of drug-lipid attractive interactions, strongly suggesting the occurrence of INH adsorption at the lipid interface \cite{xu2012predicting} and point out the relevant role of  lipid organization.  As argued by Truzzi et al. \cite{truzzi2019drugs} in PC-Chol multilamellar vesicles, drug adsorption could be originated by the presence of a certain drug-lipid affinity. Hereafter we will show that this is indeed what occurs in our formulations, corroborating this finding by an extensive characterization of the organization of lipid layer and its interaction with INH.

\section{Biophysical characterization of  liposomes and of the effect of INH interaction}
\subsection{DSC} \label{sec:DSC}
\subsubsection{Bare HSPC-DPPG liposomes}\label{subsec:DSC-bare}
\begin{figure}[htbp]
\centering
  \includegraphics[width=\linewidth]{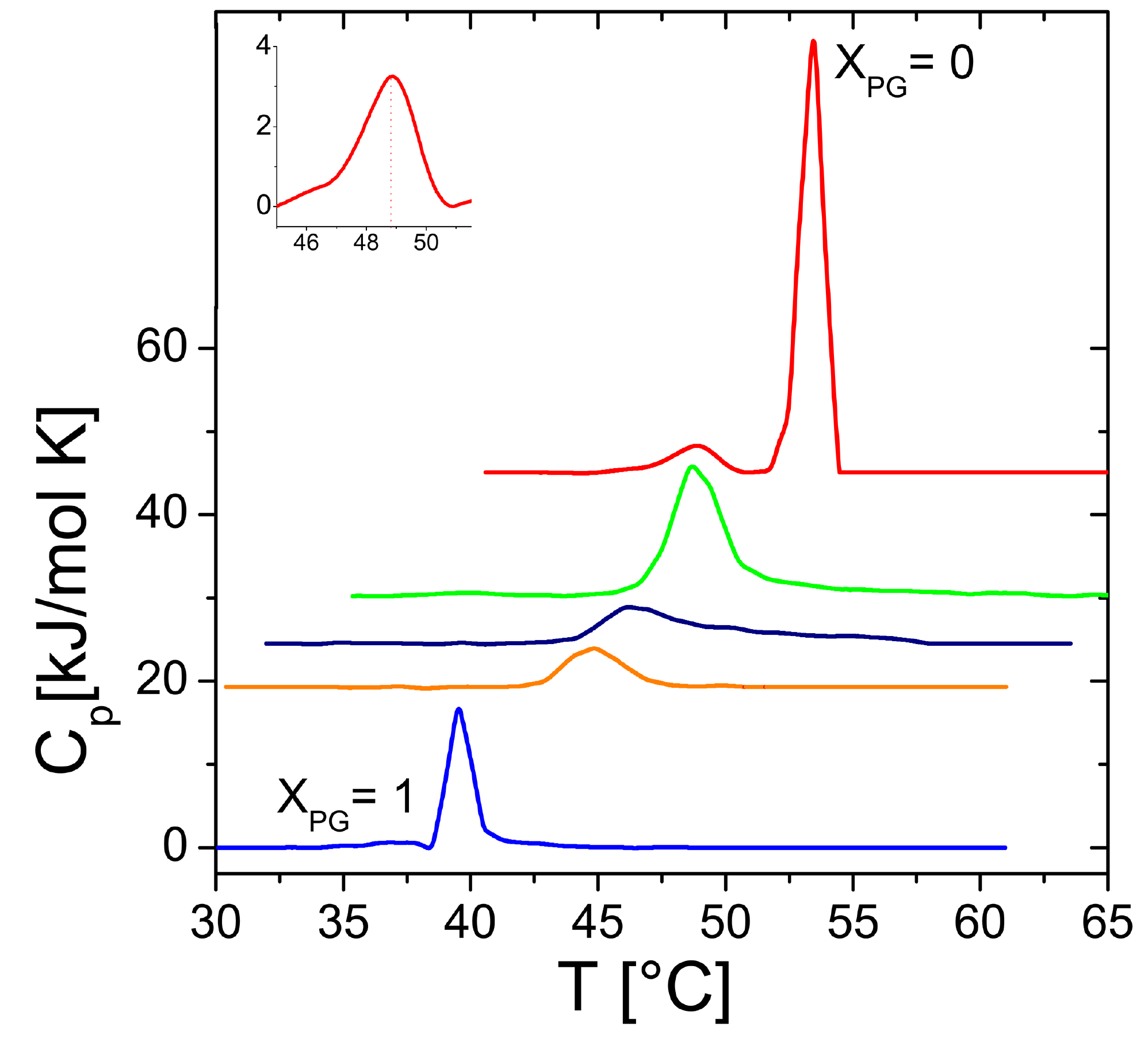}\\
  \caption{Excess molar heat capacity of multilamellar liposome suspensions (25 mg/ml) for pure DPPG  and HSPC liposomes and for different $X_{PG}$ molar ratio (increasing from bottom to top). Curves are shifted of a constant value for clarity. The inset shows the endothermic peak relative to the pretransition of HSPC. }\label{Calmulti}
\end{figure}

The DSC investigation has been performed on multilamellar liposomes to get the highest DSC signal and amplify the fine details of the local structure. Fig. \ref{Calmulti} shows the excess molar heat capacity $C_p$ obtained after baseline subtraction from the DSC thermograms of
HSPC-- DPPG multilamellar liposomes at different PG molar fraction ($X_{PG}$).

\begin{figure}[htbp]
\centering
  \includegraphics[width=0.9\linewidth]{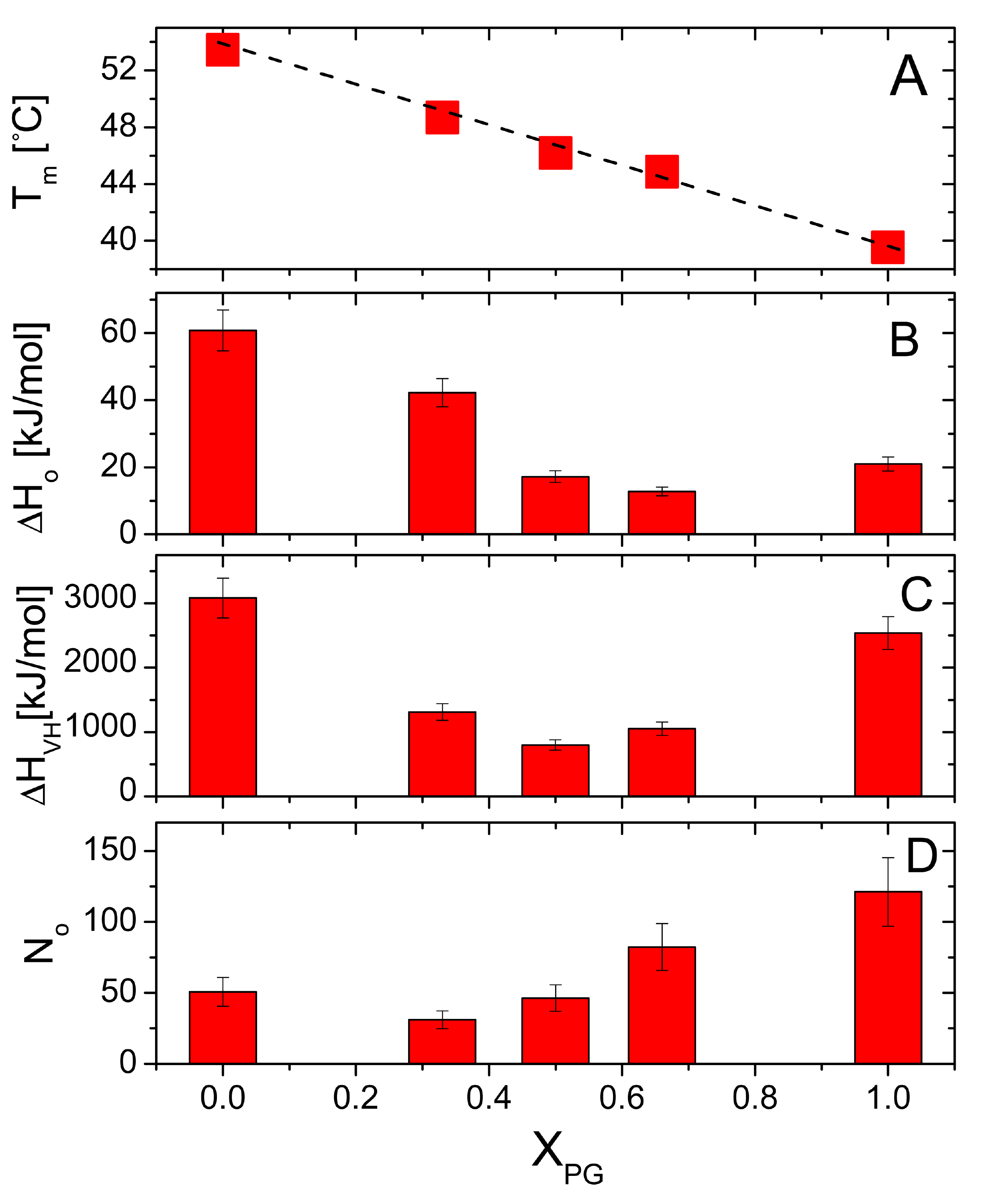}\\
  \caption{Melting temperature $T_m$ (Panel A), calorimetric enthalpy $\Delta H_0$ (Panel B), van't Hoff enthalpy $\Delta H_{VH}$ (Panel C), cooperative unit $N_0$ (Panel D) for multilamellar mixed HSPC-DPPG liposomes in function of $X_{PG}$. The dashed line in panel A is a linear fit of the data.}\label{analysmulti}
\end{figure}

HSPC liposomes exhibit two endothermic peaks corresponding to the pre- and main transition
at 48.8 $^{\circ}$C and 53.4 $^{\circ}$C, respectively. The endothermic peak progressively shifts to lower temperatures with increasing $X_{PG}$, it broadens down to $X_{PG}=0.50$ and sharpens again as the system approached pure DPPG membranes ($X_{PG}\geq 0.33$).
At the same time the pre-transition observed for pure HSPC multilamellar liposomes (magnified in the inset in Fig. \ref{Calmulti}- top panel) vanishes as the fraction of DPPG is raised. Such a pre-transition has been already reported in other aqueous environment \cite{kitayama_thermotropic_2014} in HSPC membranes and has been attributed to the formation of periodic membrane ripples. In the literature, these two transitions are usually regarded as independent events, although recent models \cite{heimburg_model_2000} suggest that both pre- and main transition are caused by chain melting.
The main endothermic transition never shows peak splitting or two detectable separate peaks, while linearly shifting to lower temperatures for increasing $X_{PG}$ (Fig. \ref{analysmulti}-A). This feature supports the full miscibility of the two lipids that melt cooperatively in the bilayers, and it is corroborated by the minimum of the peak height occurring at the equimolar condition ($X_{PG}= 0.50$), where the broadness of the process is maximum. Finally, we point out that the beginning and the end of the transition region
in the mixtures deviate from the melting temperatures of the pure compounds, confirming again that the melting of the chains of each lipid is strongly affected by the presence of the other species.

We can go further and analyze the thermograms in terms of a two-state (gel-liquid) model \cite{marsh_cooperativity_1977},
through which we determine the average size of the cooperative unit. The latter corresponds to the number of lipids passing from one state to the other simultaneously and it is given by \cite{malcolmson_dsc_1997}
\begin{equation}\label{N0}
    N_0=\frac{\Delta H_{VH}}{\Delta H_{0}} {,}
\end{equation}
where the van't Hoff enthalpy, $\Delta H_{VH}$, at the midpoint of the phase transition, $T_m$, is given by \cite{blume_biological_1991,malcolmson_dsc_1997}
\begin{equation}\label{N0}
   \Delta H_{VH}=\frac{4RT_m^2(\Delta C_p)|_{T_m}}{\Delta H_{0}} {.}
\end{equation}
Here $\Delta H_{0}$ is the calorimetric enthalpy and it is determined by integrating the DSC peak from the onset temperature, where the deviation from the baseline starts, to where the signal returns to the baseline. $(\Delta C_p)|_{T_m}$ is the peak height of the excess enthalpy. Fig. \ref{analysmulti} (B,C,D) show $\Delta H_{0}$, $\Delta H_{VH}$ and $N_0$ for the investigated samples.
The calorimetric enthalpy $\Delta H_{0}$ shown in Figure \ref{analysmulti}-B shows a weak, albeit detectable, non-monotonic behavior as a function of the DPPG fraction $X_{PG}$. We attribute the initial decrease of $\Delta H_{0}$ (from $X_{PG}=1$ to $X_{PG}=0.66$) to the effect of the increased translational entropy term due to the addition of a small amount of longer alkyl-chains in membranes mostly made of shorter chains. This tends to fluidize the membrane, increases the susceptibility of the bilayer to compositional fluctuations \cite{longo_stability_2009y} and lowers the overall amount of energy necessary to melt the membrane. As $X_{PG}$ is further decreased, the calorimetric excess enthalpy increases, as expected when the fraction of the neutral lipid (HSPC) with longest tails prevails. In this case more energy must be transferred to the suspensions in order to melt membranes in which lipid-lipid Van der Waals attractive interactions are not anymore counterbalanced by electrostatic repulsions. Fig. \ref{analysmulti}-C shows the van't Hoff enthalpy $\Delta H_{VH}$ for the same systems.
In this regard we point out two important facts: i) $\Delta H_{VH}>\Delta H_{0}$ for all $X_{PG}$, precluding the existence of multistep transitions \cite{saboury_clarification_1994} and ii) an evident minimum is present at the equimolar composition $X_{PG}=0.5$. Indeed under this condition and complete miscibility, the system reaches its maximum heterogeneity and we do expect the minimization of the transition cooperativity as the amount of mixed HSPC-DPPG contacts is the largest possible. This brings directly to the minimization of the temperature dependence of the reaction constant determining the equilibrium between lipids in the two states (melt and solid)\cite{marsh_cooperativity_1977}. Finally, figure \ref{analysmulti}-D shows the size of the cooperative units $N_0=\frac{\Delta H_{VH}}{\Delta H_{0}}$. This shows a minimum close to the equimolar composition in agreement with previously reported results for other mixed vesicles \cite{losada-perez_phase_2015}. Given our experimental uncertainty we are not able to discern possible asymmetries due to the different cooperativity of the transitions in pure one-component liposomes. 

It is important to note that, on one hand, multilamellar vesicles are the optimal system to study the thermodynamic properties of lipid arrangement since the several enclosed bilayers result in a high DSC signal. On the other hand, their large size and not controlled number of bilayers represent a serious limits in the investigation of drug-lipid interaction in a drug carrier, so that multilamellar vesicles are not anymore a suitable model for the scope of the present investigation. For a reliable modelling of the conditions characterizing the carrier, unilamellar liposomal vesicles have been considered (see SI, section 2). No detectable difference between multi- and unilamellar vesicles have been observed as far as their calorimetric behavior is concerned, corroborating once again the full miscibility of the two lipid components.

This preliminary calorimetric analysis of bare liposomes represents a dutiful premise and a necessary step summarizing the role of lipid composition for the in-depth understanding of the drug-liposome interaction that will be discussed hereafter.

\subsubsection{Interaction with Isoniazid} \label{subsec:DSC-INH}

\begin{figure}[!htbp]
\centering
  \includegraphics[width=0.9\linewidth]{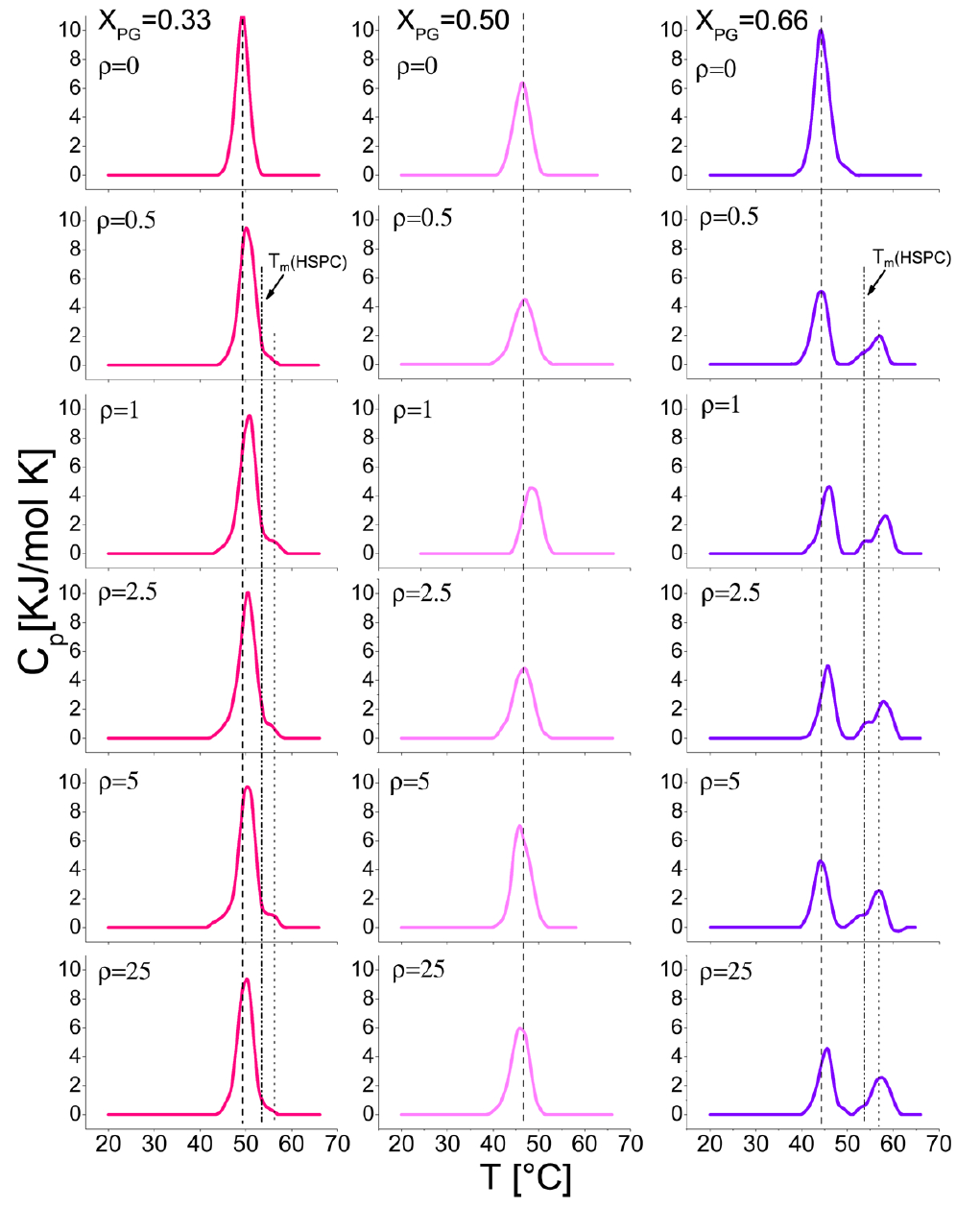}\\
  \caption{Excess molar heat capacity of unilamellar liposome suspensions (10 mg/ml) for different DPPG molar ratio $X_{PG}$ and INH/lipid molar ratio $\rho$ as indicated in the panels. Lines identifying the position of melting temperatures of peaks have been drawn as guide for eyes (dashed lines: main peak at $\rho=0$;  dashed-dot lines:  $T_m$ of pure HSPC liposomes; dot lines: secondary peak appearing at $\rho \geq 0.5$.)
  }\label{fig:calinh}
\end{figure}

Fig. \ref{fig:calinh} shows the excess molar heat capacity for the three investigated formulations of mixed liposomes and selected INH/li\-pid molar ratio $\rho$. Thermograms in absence of drug ($\rho$=0) are shown as reference in the top panels. The vertical dashed lines mark the position of $T_m$ for bare liposomes at the different compositions, pure HSPC formulations and INH-HSPC segregated domains.

First we note that the shape and the position of the main transition peak is not drastically altered by INH addition for all formulations. This confirms liposome stability at all the investigated INH concentration.

However, what is interesting and, in some respects, surprising, is the effect of INH on lipid miscibility.
For liposomes at $X_{PG}=0.5$ the thermograms do not undergo any noteworthy variation, whereas INH has a relevant effect on liposomes for the asymmetrical formulations. At $X_{PG}=0.33$ a shoulder on the right side of the main peak appears and  $X_{PG}=0.66$ this effect is even more striking. Here, in fact, the evident peak splitting occurring at any INH content is a clear signature of lipid segregation induced by drug-lipid association.

To interpret this complex behavior, one has to consider that the protocol chosen provides that INH is added to suspensions of previously formed unilamellar liposomes (as described in section \ref{sub:DSC}), thus imposing that INH molecules interact only with the external lipid layer, while the inner layer stays scarcely affected by the presence of the drug. Actually, the internal layer may be affected, in principle, by any change in local composition of the external leaf induced by the drugs. In fact, the two leaves of a lipid membrane are coupled in some way either by the interdigitation of hydrocarbon tails or through the rapid exchange of cholesterol units \cite{putzel2008phase}. However, while the former mechanism is presumably absent or negligible in our bilayers since the mismatch between the alkyl chains is very small, the latter is strictly absent.
In this situation, i.e. in presence of a weak coupling, the outer and the inner layer can show different thermodynamic phases \cite{putzel2008phase}. That is indeed what our DSC thermograms suggest for the asymmetric mixtures as a result of INH-lipid interaction.

\begin{figure*}[!htp]
\centering
  \includegraphics[width=15 cm]{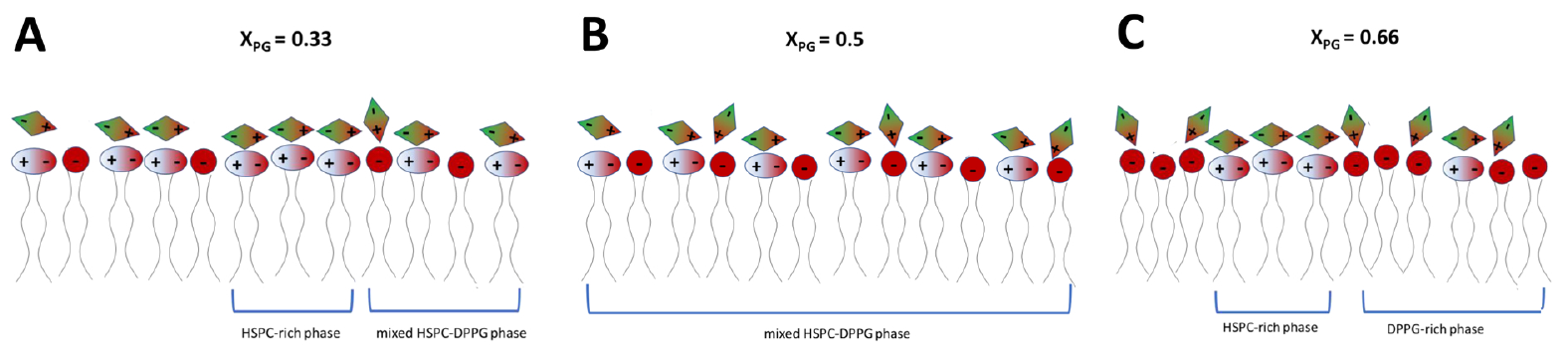}\\
  \caption{Schematic picture modelling the effect of INH addition on lipid organization. HSPC and DPPG lipids have been identified by indication of their electrostatic properties, INH molecule is drawn as a diamond. }\label{fig:INHlipid}
\end{figure*}

To interpret this complex behavior, one has to consider that the protocol chosen provides that INH is added to suspensions of previously formed unilamellar liposomes (as described in section \ref{sub:DSC}), thus imposing that INH molecules interact only with the external lipid layer, while the inner layer stays scarcely affected by the presence of the drug. Actually, the internal layer may be affected, in principle, by any change in local composition of the external leaf induced by the drugs. In fact, the two leaves of a lipid membrane are coupled in some way either by the interdigitation of hydrocarbon tails or through the rapid exchange of cholesterol units \cite{putzel2008phase}. However, while the former mechanism is presumably absent or negligible in our bilayers since the mismatch between the alkyl chains is very small, the latter is strictly absent.
In this situation, i.e. in presence of a weak coupling, the outer and the inner layer can show different thermodynamic phases \cite{putzel2008phase}. That is indeed what our DSC thermograms suggest for the asymmetric mixtures as a result of INH-lipid interaction.

On the basis of the observed DSC thermograms, in the outer lipid layer we can hypothesize the formation of super-bound INH-HSPC-rich phases with a $T_m$ higher than pure HSPC, as visible in the secondary peaks occurring at $\rho \geq 0.5$ at $X_{PG}=0.66$ (see dotted lines in Fig. \ref{fig:calinh}). At $X_{PG}=0.33$ the fraction of super-bound INH-HSPC-rich is low and the mixed HSPC-DPPG phase prevails, since a complete HSPC segregation would imply that charged DPPG molecules to get closer and closer,  this configuration being unfavored due to the high energetic and entropic penalty. In excess of DPPG ($X_{PG}=0.66$), the situation is even more complex since bare HSPC molecules segregate in a distinct phase, as evidenced by the unambiguous presence of a process centered at the melting temperature of pure HSPC that appears as a left-shoulder of the secondary peak at high temperature.  For both $X_{PG}=0.33$ and  $X_{PG}=0.66$ the segregated phases coexist with mixed HSPC-richer and DPPG-richer phases respectively, which shift to higher temperature due to the INH screening, as the main DSC peaks indicate.

It is therefore evident that the presence of INH at lipid interface and its interaction with lipids play a key role. Previous investigations  suggested the preferential surface location of INH in one-component zwitterionic DMPC or ionic DMPG liposomes \cite{rodrigues2001spectrophotometric} and hypothesized that the interaction between drug and the phosphate region of the lipid polar heads via van der Waals or hydrogen bonding can modify the lipid packing in DPPC liposomes \cite{truzzi2019drugs,marques2013isoniazid}.
At physiological condition, INH is a non-charged species \cite{pinheiro_interactions_2014}. Its local electrostatic properties and its charge density distribution have been considered relevant for its interaction with drug receptors and lipid membranes \cite{rajalakshmi2011understanding}. In particular, INH has a larger dipole moment than water \cite{bajpai2014structural}, originating from the deformation of the electronic charge distribution in the vicinity of O(1) and N(1) atoms, due to their large electronegative potential \cite{rajalakshmi2011understanding}.

In the presence of HSPC-DPPG mixed bilayers and in aqueous environments, it is then reasonable to consider that a dipole-dipole interaction between INH molecules with the zwitterionic HSPC lipids is favored. This is in line with what has been already observed for other dipolar molecules such as anesthetics \cite{kane2000interaction,cseh1999interaction}, while INH is less affine to the ionic DPPG lipid.
We speculate that this preferential attractive interaction can promote the formation of a quadrupolar INH-HSPC complexes and increases the binding energy between HSPC lipids. This mechanism can favor the segregation of the HSPC in condition of molar excess of DPPG, as suggested by the onset of a secondary peak at a temperature higher than that characterizing the melting of the bare HSPC membranes. On the other hand, the positive region of INH dipole can interact also with the anionic polar head of DPPG, thus stabilizing the DPPG-richer phase by electrostatic screening.
A simple naive scheme of this situation is sketched in Fig. \ref{fig:INHlipid}.

\begin{figure}[htbp]
  \centering
  \includegraphics[width=0.9\linewidth]{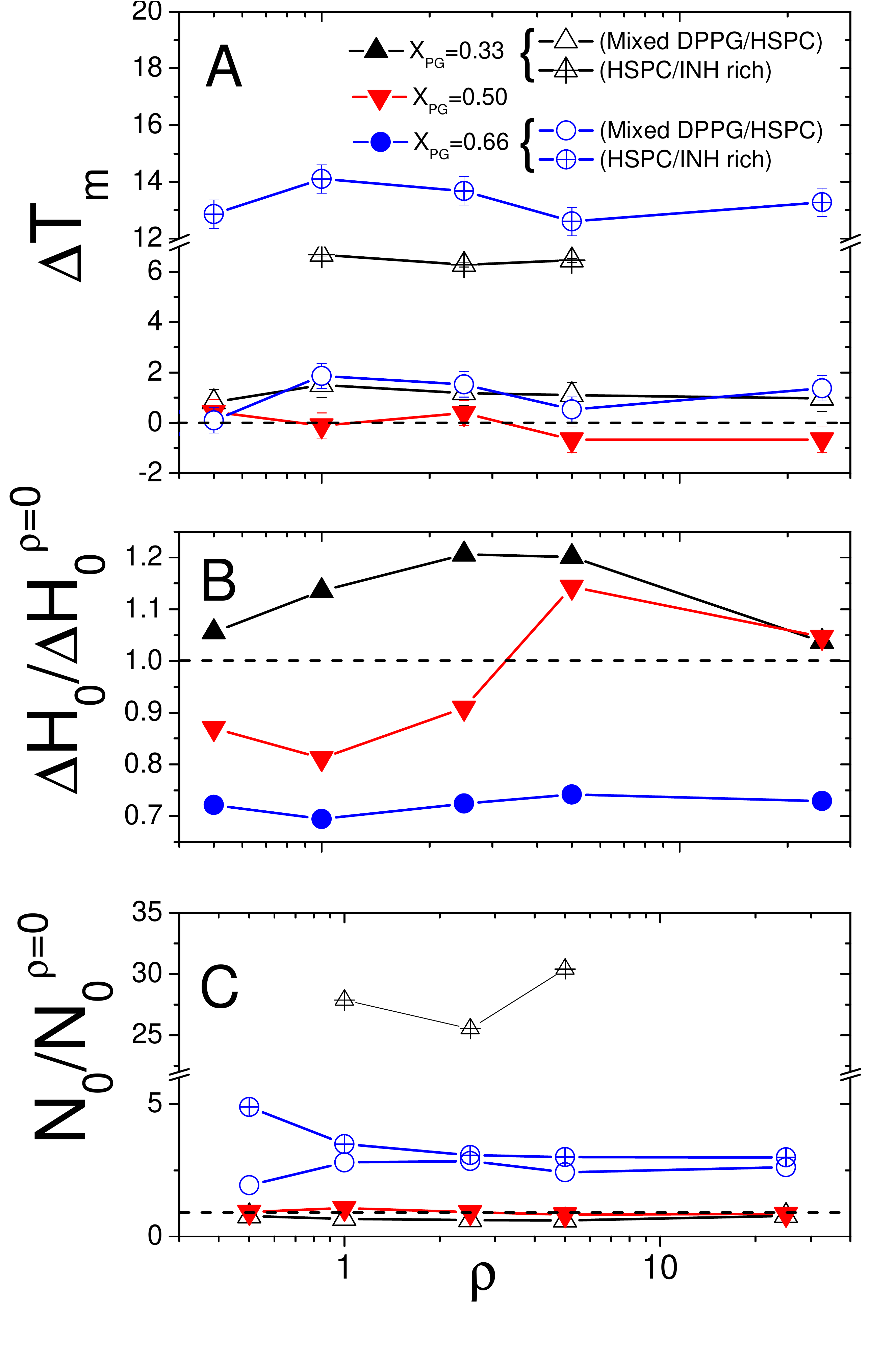}\\
  \caption{Panel A: Melting temperature shift (Panel A), normalized calorimetric enthalpy (Panel B) and normalized cooperative unit (Panel C) in function of the INH/lipid molar ratio $\rho$ at different $X_{PG}$ as indicated in panel A. The horizontal dashed lines correspond to the $\rho=0$ case.}
  \label{fig:DatiCalINH}
\end{figure}

Lipid segregation or enhanced disorder typically gives rise to an interplay between cooperativity change and shift of the melting transition temperature of the single lipids, which reflects the formation or disruption of homogeneous domains \cite{koiv_differential_1994}. For this reason, a more detailed description of the effect of INH addition can be captured by simultaneous analysis of i) the shift of the melting temperature $\Delta T_m$, ii) the variation of the calorimetric enthalpy and iii) the cooperative unit change for each endothermic transition. $\Delta T_m$ is defined as the difference $T_m^{\rho}-T_m^{\rho=0}$ between the melting temperature of lipid membranes in presence of INH and that measured for $\rho=0$ (no added INH). Analogously, the enthalpy and cooperative unit variations are calculated as the ratios between $\Delta H_0$ and $N_0$ measured in the presence of INH and the same quantities measured for bare liposomes ($\Delta H_0^{\rho=0}$ and $N_0^{\rho=0}$). Finally, to decouple endothermic peaks relative to DPPG-reach and HSPC-reach domains for $X_{PG}=0.33$ and evaluate the corresponding melting temperature and calorimetric enthalpy, we have further performed a double Gaussian fit. The results are shown in Fig. \ref{fig:DatiCalINH}.

For asymmetric mixtures, INH addition produces a positive shift of the melting temperature for all the detected processes. This suggests that primarily INH screens electrostatic repulsions between the phosphate groups decorating the membrane surfaces, lowering the effective surface charge of the liposomes and enhancing lipid local order \cite{forsyth_phase_1977}. As also observed by Pinheiro et al. \cite{pinheiro_interactions_2014} for DPPC liposomes, the positive shift of $T_m$ indicates the presence of drug at the lipid interface in the chain region (C1-C9) close to the polar heads.
Clearly, the largest increase is obtained for the HSPC-rich domains appearing in the mixtures at $X_{PG}=0.33$ and $X_{PG}=0.66$, where lipid demixing may take place \cite{giatrellis_nucleic_2011}, while for $X_{PG}=$0.5 this shift is barely detectable or even weakly negative for large $\rho$ values, suggesting that the high mixing entropy of the mixture in this case dominates over any enthalpic change due to the INH insertion. The weak, albeit detectable, negative trend observed for all the formulations in large excess of INH is presumably the signature of a local disorder induced by the INH insertion in the bilayer, that facilitates the melting transition.

At the same time we observe the calorimetric enthalpy variation decreasing progressively as the amount of DPPG increases in the mixed liposomes. This is indeed a quite remarkable result, since it unambiguously proves that compositional differences in lipid membranes determine the amount of heat needed to melt the bilayers in the presence of INH, the latter playing a pivotal role in dictating the differences between the response of lipid formulations. For the sake of clarity we remind here that the enthalpy variation for $X_{PG}=$0.66 and $X_{PG}=$0.33 refers to the whole melting process and not only to one of the two endothermic transitions observed for such mixtures.

We now try to detail even more the description of the calorimetric behavior of the suspensions.
The mixed HSPC-DPPG phases in two asymmetric mixtures show a quite specular behavior, both in terms of calorimetric enthalpy variation, (Figure \ref{fig:DatiCalINH}-B), and normalized cooperative unit (Fig.\ref{fig:DatiCalINH}-C).
At $X_{PG}=0.33$ the INH addition causes an overall increase of the energy necessary to bring all the lipids in the melt state and a weak reduction of the cooperative unit, suggesting that the screening of the surface charge increases the average energy barrier between the melt and the solid state of the lipids, bringing at the same time the cooperative unit closer to the one obtained for pure HSPC membranes, since the charge defects introduced by the phosphate groups of the DPPG are now presumably compensated by INH adsorption. By contrast, a large cooperative unit variation characterizes the HSPC-rich domains, suggesting that lipid-lipid correlations are enhanced by the INH-HSPC coupling. We refer the reader to section \ref{monolayers} for a further discussion of the occurrence of lipid demixing in asymmetric HSPC-DPPG formulations.

For $X_{PG}=0.66$ the decrease of the calorimetric enthalpy with respect to the bare liposomes is accompanied by an increase of the cooperativity of both the mixed HSPC-DPPG domains, which is generally attributed to a higher solvation and reduced charge of the phospholipid head groups \cite{nunes2011lipid}, and HSPC-rich domains.
In the latter case the large cooperativity of the melting process is strictly related to demixing. As previously discussed, INH addition, is able to strongly destabilize the lipid mixture, inducing lipid segregation and phase separation also for a relatively small amount of the drug. A higher affinity between INH and HSPC, for which dipole-dipole short range interactions are large compared to the dipole-monopole interaction between INH and DPPG rationalizes this result.
It's worth stressing again that unbalanced interactions between lipids and an external agent (INH here) are very important in determining the lipid local order within the membrane, as they typically favor segregation of one of the two species, as also reported for DNA-cationic liposome complexes \cite{koiv_differential_1994,giatrellis_nucleic_2011, iglic_advances_2014,harries_structure_1998, bruinsma_long-range_1998}.
All in all, in excess of PG headgroups INH addition enhances the formation of DPPG-rich (mixed) and HSPC-rich (pure) domains, both characterized by high cooperativity, since in HSPC-rich domains lipids are more strongly bound and, at the same time, INH screens residual charges borne by PG headgroups.

It is also interesting to note that the two components of the demixed membranes respond differently to an increase of INH content in the $X_{PG}=0.66$ formulation. On the one hand, the cooperativity of the transition of DPPG-rich domains first increases for $\rho \lesssim 2.5$,  presumably due to the electrostatic coupling with INH molecules, and then weakly decreases for larger values of $\rho$, probably due to INH insertion in the bilayer. On the other hand, the values of the cooperative unit of HSPC-rich domains decreases progressively for increasing values of $\rho$, showing that the a large amount of INH molecules in the bilayer mainly degrades molecular correlation and cooperativity in zwitterionic domains.

Actually, as observed also for nucleic acid-lipid complexes \cite{koiv_differential_1994,giatrellis_nucleic_2011}, INH molecules could affect more intimately the lipid bilayer structure than simply enhancing electrostatic screening by molecular insertion. This may favor disorder and nucleation of defects and facilitate the transition to the melt state of the mixed bilayer.

Finally we note that for $\rho \gtrsim 2.5$ and for $X_{PG}=0.5$ local disorder is enhanced, as reflected by a decreased melting temperature for high INH content. The interpretation of the data for the symmetric mixture is definitely more intricate as cooperativity stays basically unaffected by INH addition, while the calorimetric enthalpy scatters and suggest a subtle balance between the onset of demixing, electrostatic screening and INH insertion within the membrane.
This aspect deserves by all means a more detailed investigation through techniques probing more directly the membrane structure, including more refined calorimetric characterization, and will be the subject of a future publication.

\subsection{Static light scattering}

By taking advantage of Static light scattering (SLS) we measured the time-averaged scattered intensity at different scattering vectors $q$ (see section \ref{MatMeth}), and hence the form factor and the gyration radius $R_g$ of the liposomes. At the same time, it is interesting to investigate the intensity at one fixed scattering angle (here $I_{90}$ measured at $\theta=90^{\circ}$) as a function of temperature, since a change of the refractive index of the vesicles induced by the membrane melting transition affects this observable \cite{michel_determination_2006}. We have then characterized the melting transition of the lipid bilayers via light scattering technique (results are reported in SI). Finally, we also note that the melting transition of all the liposomes investigated here gives rise, as expected, to a thinning of the lipid bilayers (see SI).

\begin{figure}[htbp]
  \centering
  \includegraphics[width=6.5cm]{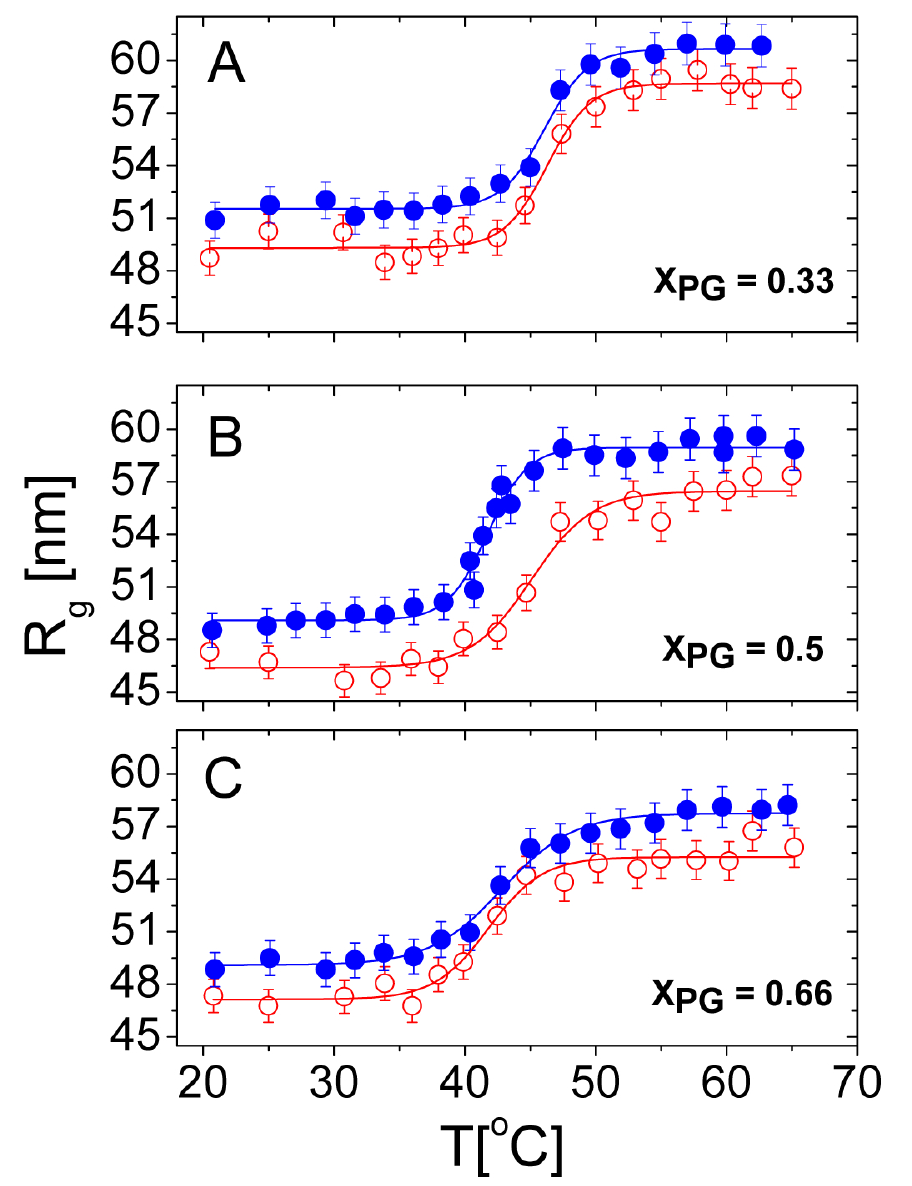}\\
  \caption{Gyration radii of the unilamellar liposomes for bare liposomes ($\rho=0$, empty symbols) and in the presence  of INH ($\rho=5$, full symbols) for $X_{PG}=0.33$ (Panel A), $X_{PG}=0.50$ (Panel B), $X_{PG}=0.66$ (Panel C). }\label{rgINH}
\end{figure}

The gyration radii of all mixed liposomes in absence of INH $\rho=0$) and at a selected drug concentration ($\rho=5$) is shown in Fig. \ref{rgINH}.  INH addition gives rise to an increase of $R_g$ of the liposomes for all the $X_{PG}$ investigated and at all temperatures. This rules out 1) the simple electrostatic screening effect due to the INH localization within a diffuse layer around the liposomes that would facilitate the lipid compactness within the bilayer and hence an overall decreases of the liposome size, and 2) an osmotic effect due to the excess of solutes outside the liposomes giving rise to the partial evacuation of water from the interior of the liposomes \cite{sabin_size_2006} and hence to their shrinkage. On the other hand, the increased size suggests that INH does accumulate on liposome surface possibly penetrating (even partially) in the bilayer.
As a matter of fact both superficial adsorption and/or partial insertion of INH within the bilayer would give an increase of the liposome size.

By fitting the $R_g(T)$ with a Boltzmann-type equation (eq. \ref{Boltzmann2}) we have extracted the optical transition temperature  $T_c^{opt}$ for bare liposomes and in the presence of INH ($\rho=5$).
Consistently with the DSC results at $\rho=5$, we observe a net decrease of the transition temperature with respect of the bare liposomes only in the case of equimolar mixtures ($X_{PG}=0.5$) (see table  \ref{tab:tablerho5}), while for the other two mixtures the shift is positive ($X_{PG}=0.66$) or not detectable ($X_{PG}=0.33$) (see table  \ref{tab:tablerho5}).

\begin{table}
\footnotesize
\begin{center}
\centering
\caption{Melting temperatures ($T_c^{opt}$) and gyration radii ($R_s$,$R_m$) below and above the melting transition of unilamellar liposomes for the three $X_{PG}$ employed in this work  for bare liposomes ($\rho=$0) and at a fixed INH/lipid molar ratio ($\rho=$5). All values are obtained by a non-linear fit of $R_g(T)$ via equation \ref{Boltzmann2}. \label{tab:tablerho5}}
\begin{tabular}{ccccc}
$X_{PG}$ &$\rho$& $R_s$ [nm]& $R_m$ [nm] & $T_c^{opt}$\\
\hline
0.33&0 & 49.3 $\pm$ 0.5 & 58.6 $\pm$ 0.5& 46.1 $\pm$ 0.5\\
0.33&5& 51.5 $\pm$ 0.2 & 60.6 $\pm$ 0.4 & 46.1 $\pm$ 0.3\\
&&&&\\
0.50&0 & 46.4 $\pm$ 0.5 & 56.4 $\pm$ 0.5& 43.9 $\pm$ 1.0\\
0.50&5&  49.1 $\pm$ 0.2 & 58.9 $\pm$ 0.4 & 38.7 $\pm$ 0.2\\
&&&&\\
0.66&0& 47.1 $\pm$ 0.7 & 55.3 $\pm$ 0.5& 40.5 $\pm$ 0.9\\
0.66&5& 49.1 $\pm$ 0.2 & 57.7 $\pm$ 0.4 & 42.8 $\pm$ 0.5\\
\end{tabular}
\end{center}
\end{table}

Once again, all the above described results corroborates the scenario where INH molecules strongly interact with mixed HSPC-DPPG bilayer and affect its internal structure by binding preferentially with one type of lipid rather than with both the components of the bilayer at the same extent.

\subsection{Monolayers}\label{monolayers}
To further clarify the interaction of INH with  HSPC and DPPG and unveil the very nature of the interaction of INH with both the lipids
we have studied the isotherms of mixed HSPC-DPPG and pure (one-component) DPPG and HSPC lipid monolayers at air-water interface in the presence of INH. The drug has been dissolved in ethanol and injected in the subphase under the monolayers in the liquid-expanded phase. The so-formed layers have been then compressed. The INH insertion in this precise condition allows to observe the maximum extent of drug-lipid interaction, since the low lipid density of the monolayer facilitates drug-lipid association  \cite{chimote2008evaluation}.

\begin{figure}[htbp]
  \includegraphics[width=\linewidth]{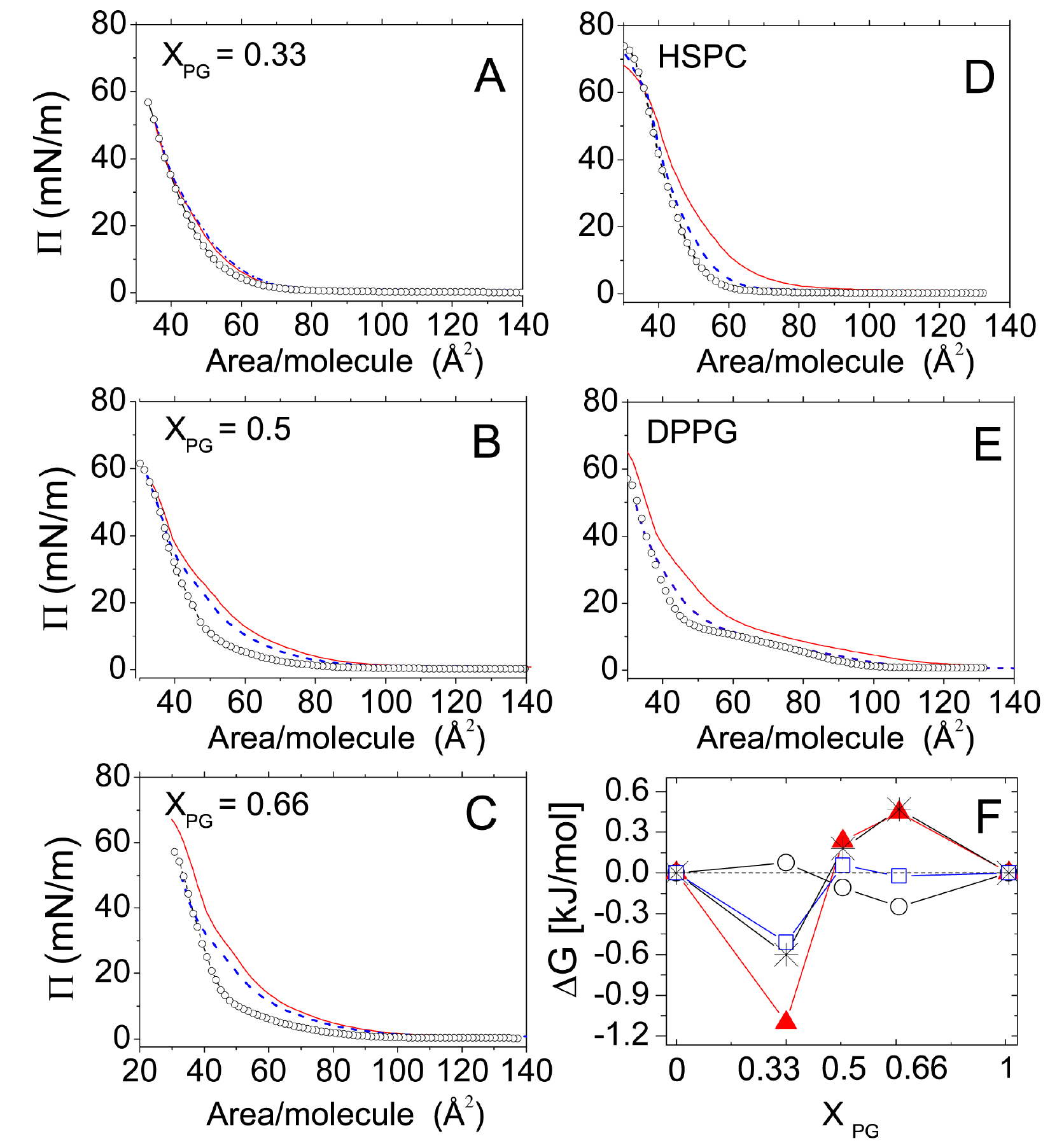}\\
  \caption{Surface pressure isotherms of mixed HSPC-DPPG monolayers (A-B-C) and pure lipids (D-E) on Hepes subphase  ($\bigcirc$) and after injection of 200 $\mu$l of ethanolic INH or pure ethanol (full and dashed line, respectively).  Panel E shows the excess free energy of mixing $\Delta G$  calculated for mixed HSPC-DPPG monolayers on pure hepes subphase $\bigcirc$) and  after injection of 200 $\mu$l of ethanolic INH solution ($\blacktriangle$) or pure ethanol ($\square$), with the calculated difference ($\ast= \blacktriangle- \square$) to determine the effect of only INH.  $\Delta G$ is calculated at $\Pi$=35 mN/m. }\label{Fig:ISO}
\end{figure}

Preliminarily, we have tuned the nominal INH/lipid molar ratio $\rho$ by injecting increasing volumes of INH, dissolved in ethanol at fixed concentration (see SI). Isotherms obtained after injection of 200 $\mu$l of INH solution  at concentration 0.013 mM, i.e. the maximal amount used in monolayer experiments,  are shown in Fig. \ref{Fig:ISO}, for mixed  (panels A-B-C) and one-component monolayers (panels D-E). To evaluate the net effect of INH,  isotherm obtained after the injection of the same volume of pure ethanol are shown for comparison. Ethanol is an amphiphilic surface-active compound and it may interact with the lipid monolayers by insertion at the interface \cite{wilson_adsorption_1997}. Injection of INH or ethanol causes a shift to larger area per molecules, at a given surface pressure. This is evident by comparing the different curves with those obtained for bare monolayers, i.e. in the absence of injection (Panel A-E of Fig. \ref{Fig:ISO}). This effect is commonly attributed to the adsorption of the injected molecules at the interface and is enhanced by the interaction between the drug and lipid layers, as claimed by Chimote et al. \cite{chimote2008evaluation} and by Marques et al. \cite{marques2013isoniazid}, who suggested an intercalation of INH close to the polar head of DMPC liposomes. Since here INH is in large excess with respect to number of lipids, its insertion may occur, followed by a lipid realignment upon compression due to the varying balance between the lateral attractive/repulsive forces between lipids upon increasing their packing.

In fact, if the amount of molecules inserted in the lipid film does not change during compression, an increase of the molecular areas in the condensed phase due to the additional space needed for INH molecules is expected. This 'additional space' cannot be neglected, and indicates that the drug is located in some extent in the lipid layer.
Here we observe that $\Pi$-A isotherms of DPPG and HSPC in the condensed state have an almost-parallel course in the presence and in the absence of INH. In mixed monolayers this effect decreases upon decreasing the DPPG content. For $X_{PG}=0.33$, the isotherm recorded in the presence of INH converges towards the isotherms of the bare monolayer at high surface pressure. This could be attributed to a 'squeezing out' of drug during compression or to a peculiar rearrangement between the drug and the lipids, which minimizes the overall molecular hindrance.

Fig. \ref{Fig:ISO}-F shows the excess free energy $\Delta G$ for mixed HSPC-DPPG monolayers in the presence and the absence of INH and ethanol. The values are calculated at $\Pi= 35 mN/m$, i.e the pressure correspondent to the packing density of lipid bilayers \cite{marsh_lateral_1996,dahim_physical_2002}.
Mixed HSPC-DPPG monolayers ($\circ$) on pure Hepes subphase show an almost ideal-behavior with $\Delta G$ approaching 0 at low PG content and a slightly higher negative deviation for $X_{PG}= 0.66$, indicative of attractive interactions between the two lipids which can stabilize the mixed film. This confirms the results obtained by DSC indicating a full miscibility of the two lipids  (see section \ref{sec:DSC})
and it is in agreement with simulations \cite{longo_stability_2009y}, showing that demixing does not occur when the length mismatch between the alkyl chains of the two lipids in the mixture is lower than 6 carbons.

The addition of a polar compound as ethanol and INH ($\square$, $\blacktriangle$, respectively) causes deviations from the ideal behavior, indicating the onset of non-negligible interactions between the solutes added in the subphase and the lipid film.
The large negative deviations at $X_{PG}= 0.33$ in the presence of INH indicate that HSPC-DPPG bonds are preferred with respect to HSPC-HSPC and DPPG-DPPG ones. It's worth remarking here that this is not in contrast with the onset of super-bound INH-HSPC states observed in DSC thermograms for this lipid formulation since in large excess of HSPC the presence of such states is statistically favored even without lipid demixing.

Conversely, beyond equimolarity, the positive deviations of $\Delta G$ indicate that interaction between lipids of the same kind are favored and may cause lipid segregation.
The latter indeed can be induced in the presence of positive values of $\Delta G$ of the order of $K_B T$ at T=298 K, as in our system, in line with Monte Carlo simulations of binary lipid mixtures \cite{heberle2011phase}.
In agreement with DSC results, this finding indicates that INH-lipid interaction can modify the miscibility of HSPC and DPPG in a complex interplay with lipid composition.

We have already discussed in section \ref{subsec:DSC-INH} how the attractive interaction between the dipoles of INH and HSPC can promote the formation of quadrupolar INH-HSPC complexes and increases the binding energy between HSPC lipids. However it's worth stressing once again that this mechanism can favor the segregation of a HSPC-rich phase in excess of DPPG, as indicated by the positive values of $\Delta G$ at $X_{PG}=0.66$. While the condensation of the relatively few HSPC lipids 'diluted' in a enriched DPPG matrix is entropically  and energetically costly in the absence of INH, the addition of INH is able to modify this scenario via the formation of attractive complexes and screening of electrostatic repulsion between DPPG molecules. Conversely, in excess of HSPC, the addition of INH does not cause lipid demixing because the condensation of the DPPG molecules, now acting as repulsive 'defects', would always imply an high energetic and entropic penalty. In this condition and in excess of INH, the negative value of  $\Delta G$ suggests a stabilization of the mixed film by screening the electrostatic repulsions between DPPG molecules.

All in all, these findings corroborate the scenario in which the strong lipid demixing observed in unilamellar liposomes with $X_{PG}= 0.66$ and in the presence of INH, is driven by the modification of lipid-lipid interaction due to INH-HSPC binding.

\section{Conclusion}
We have investigated the encapsulation of the antitubercular drug isoniazid (INH)
in charged unilamellar vesicles composed by mixtures of zwitterionic HSPC and anionic DPPG lipids, and its interaction with the lipid bilayers. For the first time the amount of drug encapsulated in the vesicles, determined by UV spectroscopy, has been compared with the one expected from geometrical arguments and based on the determination of the liposome volume fraction by 'Laser transmission Spectroscopy' (LTS) technique. We found that the encapsulation of INH is much larger than the expected one, showing that drug-lipid interaction is relevant. Such a result represents indeed a crucial result of the present work and has motivated a further deep investigation of drug-lipid interaction by calorimetry, static light scattering and Langmuir monolayer technique.

INH can accumulate at the lipid interface, as indicated by the systematic $\sim$2-nm increase of the gyration radius $R_g$ of liposomes in the presence of INH, that further modifies lipid miscibility in a complex interplay between electrostatic screening, entropy, lipid-lipid and drug-lipid interaction, as shown by calorimetry. Surprisingly, we found that INH can induce lipid segregation in asymmetric mixtures which gives rise to a clear phase separation at excess of the anionic species (DPPG). Conversely, at excess of HSPC and at equimolar composition,  the screening effect of INH prevails and the lipid layer remaining fully miscible as in the absence of the drug, with the maximum heterogeneity observed at the equimolar composition. The results obtained on HSPC-DPPG Langmuir monolayers confirmed the accumulation of drug at the interface and the phase separation at DPPG excess,  pointed out the modification of the lipid packing due to INH insertion. At the equimolar composition the maximal heterogeneity of the lipid layer occurs and INH insertion in the bilayer could be favored, thus explaining the slightly larger value of entrapment ratio found for INH-loaded liposomes.

Since INH is a small dipolar molecule with the amine group protruding out of the molecular plane of the piridine ring \cite{rajalakshmi2011understanding,saikia2013density}, its peculiar structure and electronic configuration could favor the electrostatic interaction with zwitterionic lipids and its insertion in the bilayer.
In a naive picture of the INH-lipid interaction, it can be speculated that, thanks to its dipolar nature at physiological pH, INH  is more affine to the zwitterionic HSPC than DPPG and can form a quadrupolar complex which increases the binding energy between HSPC lipids. The condensation of these complexes and the segregation of a HSPC-rich phase occurs mainly at DPPG excess, where the entropic penalty due to lipid segregation can be counterbalanced by the strong enthalpy gain obtained by bringing together quadrupolar INH-HSPC complexes.

At the best of our knowledge, our investigation represents the first piece of evidence on the effect of INH on lipid organization in charged PC-based liposomes to be employed as anti-TB nanocarrier for pulmonary delivery.
While previous works dealing with uncharged lipid bilayers have shown the crucial role of bilayer composition for targeting the biological membranes and understanding the mechanism of action of INH, a comprehensive investigation on the interaction between this drug and charged liposomal nanocarriers is lacking. Only recently, a small-angle neutron-scattering (SANS) investigation focused on the structure of neutral PC-Chol multilamellar vesicles loaded with isoniazid and rifampicin and hypothesized an affinity between INH and PC \cite{truzzi2019drugs}.

Our work points out the importance of the investigation of the drug-lipid interface to improve the design of a nanocarrier.
The control of the transport of materials across the bilayer, i.e. the release of entrapped cargo from liposomes, is a critical element needed to harness the potential of lipid-based vesicular carriers. A simpler mean to exert control over efflux in synthetic liposomes involves the knowledge of lipid structure and lipid/drug interaction dictating self-assembly and permeability properties \cite{lou2020strategies}. Furthermore, the ability to understand the modification induced by drug/lipid interaction could help in finding strategies to modulate bilayer stability and semi-permeable properties and to stabilize the liposome bilayer during circulation \cite{pinheiro2019antibiotic, le2020moxifloxacin}. A deeper understanding of drug-bilayer interactions may lead to development of safer and more efficient drugs and drug delivery systems.

\section{Acknowledgments}
This research was funded by Phospholipid Research Center (Grant n. FBO-2017-051/1-1) and supported by Lipoid.
F.S and S. T. acknowledge support from Torno Subito projects of Lazio Adisu-Regione Lazio;
S. S. thanks S. Casciardi for TEM microscopy and C. Bombelli and F. Ceccacci for use of Minitrough and for scientific discussions.

\bibliography{bibliography}

\end{document}


\begin{frontmatter}

\title{
Influence of drug/lipid interaction on the entrapment efficiency  of isoniazid in liposomes for antitubercular therapy: a multi-faced investigation.\\
\textbf{Supplementary information} }

\author{Francesca Sciolla}
\address{Dipartimento di Fisica La Sapienza Universit\`{a} di Roma and CNR-ISC UOS Roma, Piazzale A. Moro 2, I-00185 - Rome (Italy)}
\author{Domenico Truzzolillo, Edouard Chauveau}
\address{Laboratoire Charles Coulomb (L2C), UMR 5221 CNRS-Universit\'{e} de Montpellier,
4 F-34095 Montpellier - France}
\author{Silvia Trabalzini}
\address{Dipartimento di Chimica e Tecnologie farmaceutiche, Universit\`{a} di Roma, Piazzale A. Moro 5, I-00185 - Rome (Italy)}
\author{Luisa Di Marzio}
\address{Dipartimento di Farmacia, Universit\`{a} G.d'Annunzio, Via dei Vestini, 66100 - Chieti, Italy}
\author{Maria Carafa, Carlotta Marianecci}
\address{Dipartimento di Chimica e Tecnologie farmaceutiche La Sapienza Universit\`{a} di Roma, Piazzale A. Moro 2, I-00185 - Rome (Italy)}
\author{Angelo Sarra}
\author{Federico Bordi}
\address{Dipartimento di Fisica La Sapienza Universit\`{a} di Roma and CNR-ISC UOS Roma, Piazzale A. Moro 2, I-00185 - Rome (Italy)}
\author{Simona Sennato}
\address{Dipartimento di Fisica La Sapienza Universit\`{a} di Roma and CNR-ISC UOS Roma, Piazzale A. Moro 2, I-00185 - Rome (Italy)}

\end{frontmatter}

\section{Encapsulation of INH:preliminary results}

A set of experiments has been performed to determine the best condition for encapsulation of INH with a INH/lipid molar ratio $\rho$ ranging from 1 to 50. The results are reported in Table \ref{tab:encap1}.
For all the formulations $X_{PG}=0.33,0.5,0.66$, the highest values of encapsulation have been found at intermediate INH/lipid ratio.

\begin{table}
\begin{center}
\caption{Encapsulation of INH for liposomes at different $X_{PG}$ and INH/lipid molar ratio $\rho$. $E.E.\%$ and $D.L.\%$ are the encapsulation efficiency. They have been calculated according to eq. 2 in the main text (section 2.7). In the case marked by the asterisk ($\ast$) the UV determination is not reliable because it was  close to the sensitivity limit of the instrument.}
\label{tab:encap1}

\begin{tabular}{cccc}
$X_{PG}$ & $\rho$  &  E.E. \%    \\
\hline
0.33     & 1     &      2.1 $\pm$ 0.2         \\
0.33     & 10     &     2.4 $\pm$ 0.2          \\
0.33     & 50     &     0.3 $\pm$ 0.02             \\
&&&\\
0.50     & 1     &        1.6 $\pm$ 0.3         \\
0.50     & 10   &              1.7 $\pm$  0.2              \\
0.50     & 50  &            0.3 $\pm$ 2        \\
&&&\\
0.66    & 1   &             $\sim$0 $*$    \\
0.66    & 10   &              1.1  $\pm$   0.1          \\
0.66     & 50   &             0.3$\pm$ 2           \\
\end{tabular}
\end{center}
\end{table}

\section{DSC investigation of unilamellar liposomes}

\begin{figure}[htbp]
 \includegraphics[width=12cm]{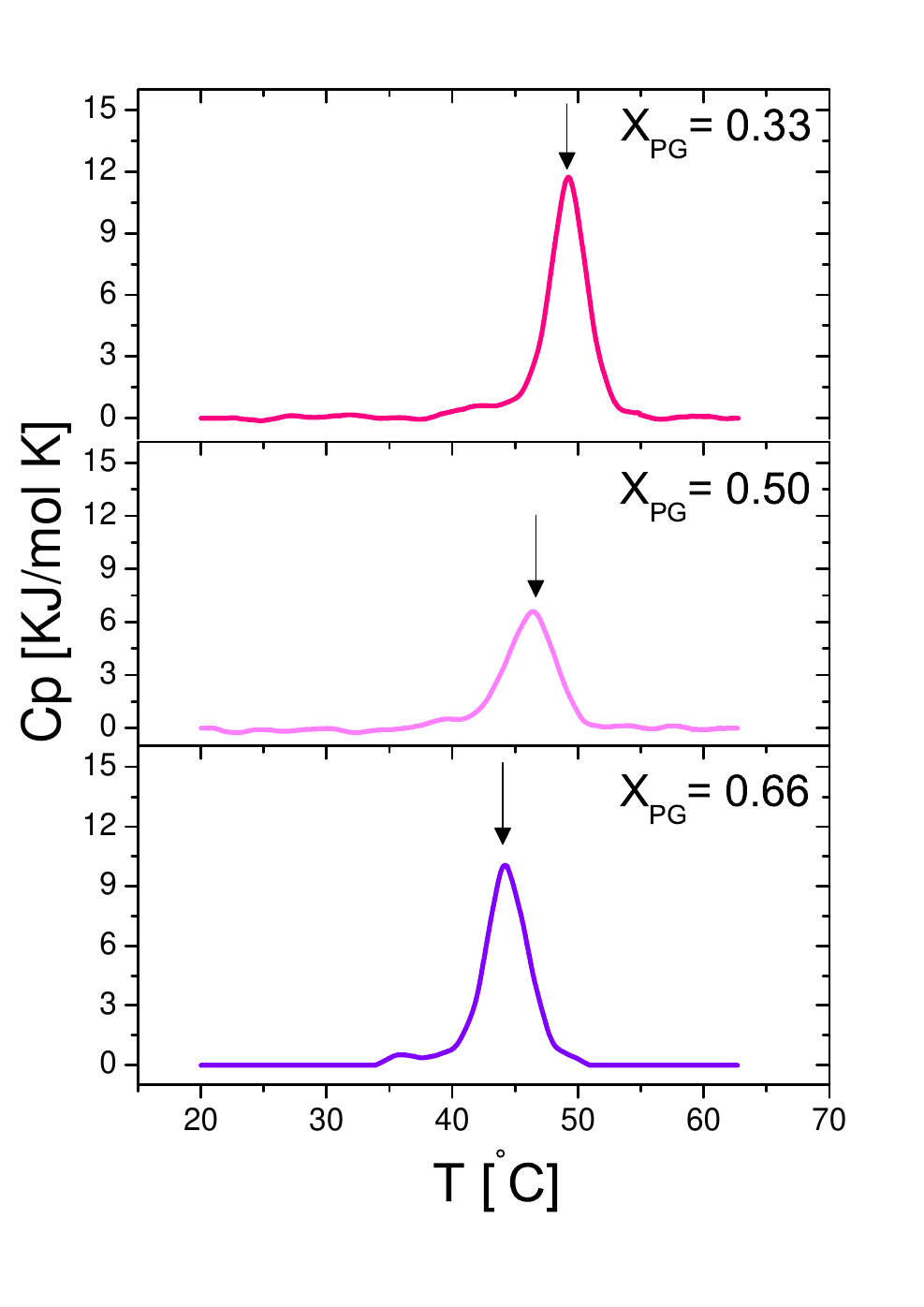}\\
  \caption{Excess molar heat capacity of unilamellar liposome suspensions (10 mg/ml) for different DPPG molar ratio $X_{PG}$ as indicated in the panels. Vertical arrows point at the peak position for each DPPG ratio.}\label{Caluni}
\end{figure}

Here we report on the DSC investigation on unilamellar liposomes.
We shortly discuss the results by comparing them with those obtained for multilamellar liposomes reported in section 4.1.1, hence isolating the role of the radius of curvature on the melting transition of the membranes.
While no significant difference is found for what concerns the transition temperature (Figure \ref{Caluni} and table \ref{tab:AnalisiCalUni}),
for all the lipid formulations ($X_{PG}=0.33, 0.5, 0.66$) unilamellar liposomes are characterized by larger enthalpies with respect to their multilamellar analogs.  This difference is barely detectable $X_{PG}=$0.33 and becomes increasingly large as $X_{PG}$ increases. Such an effect is in line with the confinement due to the small curvature radius of unilamellar vesicles, since it brings on average the lipids more in contact and increases the amount of energy needed to disorder completely the alkyl-chains within the bilayers \cite{marsh_cooperativity_1977,conti_physical_1985}. We speculate that this reduced difference between unilamellar and multilamellar HSPC-rich liposomes may depend on the larger flexibility of the HSPC-rich membranes as they are less charged. Indeed membrane flexibility is lowered by both increasing surface charge density $\sigma_c$ and thickness \cite{claessens_bending_2007}.
In our case, by decreasing the DPPG fraction, liposomes are progressively less charged and we expect to have a quite important decrease of the mean bending modulus $k_c$ of the bilayer, since the latter is expected to scale as $k_c\propto\sigma_c^2\sim (X_{PG})^2$ \cite{winterhalter_effect_1988-1,lekkerkerker_contribution_1989}.

By contrast, $\Delta H_{VH}$ does not show the same difference between multilamellar and unilamellar liposomes and no particular trend in function of $X_{PG}$ is detected. This finding can be interpreted in the light of the obtained size of the cooperative unit, that turns out to be systematically lower in unilamellar vesicles, with the largest difference measured for the most charged mixed liposomes ($X_{PG}$=0.66). Such a result suggests that small radii of curvature suppress cooperative transitions \cite{marsh_cooperativity_1977,conti_physical_1985} and that this occurs more pronouncedly in charged lipid bilayers.

\begin{table}
\begin{center}
\caption{Melting temperature $T_m$, calorimetric enthalpy $\Delta H_0$, Van't Hoff enthalpy $\Delta H_{VH}$, and cooperative unit $N_0$ for unilamellar mixed DPPG-HSPC liposomes in function of $X_{PG}$.}
\label{tab:AnalisiCalUni}
\begin{tabular}{c c c c c }
$X_{PG}$ &  $T_m$  [C$^o$]&  $\Delta H_0$ [kJ/mol]  & $\Delta H_{VH}$ [kJ/mol] & $N_0$ \\
\hline
0.33     &   44.2 $\pm$ 0.2 &      44.1 $\pm$ 4.4    &  770$\pm$ 77  &  17.4 $\pm$   3.5 \\
0.50     &   46.4$\pm$ 0.2&       32.0$\pm$3.2    &  699$ \pm $ 70    &  21.8$\pm$   4.3  \\
0.66     &   49.2 $\pm$ 0.2&     45.3 $\pm$ 4.5  & 895$\pm$ 89   & 19.7$\pm$3.9  \\
\end{tabular}
\end{center}
\end{table}

\section{SLS investigation of bare unilamellar liposomes}
By static light scattering it is interesting to investigate the intensity at one fixed scattering angle (here $I_{90}$ measured at $\theta=90^{\circ}$) as a function of temperature, since a change of the refractive index of the vesicles induced by the membrane melting transition affects this observable \cite{michel_determination_2006}.
We have then characterized the melting transition of the lipid bilayers via light scattering technique, which represents a reliable, simple, and reproducible way to determine important biophysical parameters of the membranes by investigating the mean count rate (average number of photons detected per second) \cite{michel_determination_2006}, complementary to calorimetric investigation.

\begin{figure}[htbp]
  \includegraphics[width=\linewidth]{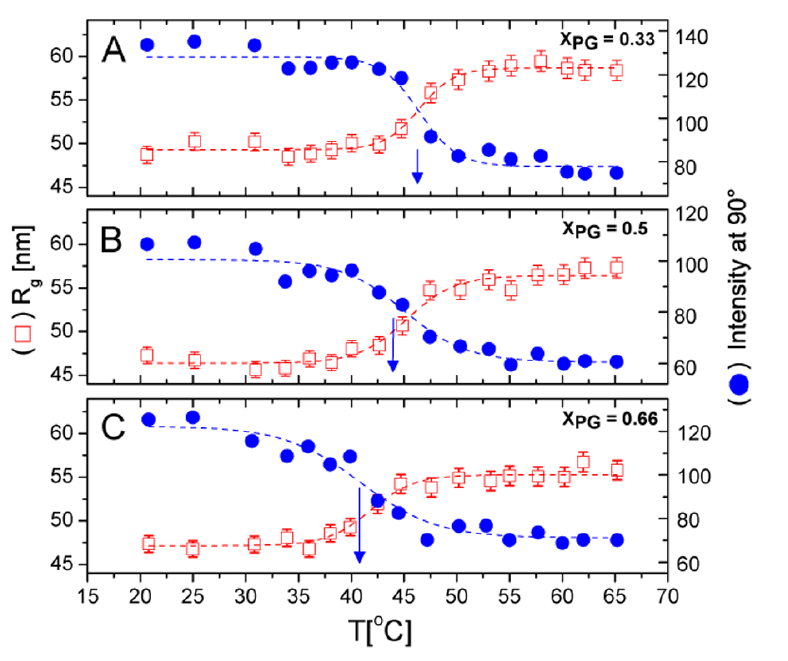}\\
  \caption{Scattered light intensity at $\theta$=90 ${^\circ}$C (Blue) and gyration radius $R_g$ (Red) for mixed HSPC-DPPG liposomes at different DPPG molar ratio $X_{PG}$ as shown in the panels. Lines are the best fits obtained by using the Boltzmann-type equations \ref{Boltzmann} and \ref{Boltzmann2}, arrows mark the melting temperatures $T_c^{opt}$}\label{Irg}
\end{figure}

Fig. \ref{Irg} shows  $R_g$ as a function of temperature of the mixed unilamellar liposomes and the relative scattered light intensity measured at fixed scattering angle ($\theta=90^{\circ}$) for the three investigated $X_{PG}$. Both $R_g$ and $I_{90}$ have been measured by illuminating samples with the same lipid mass fraction (0.2 mg/ml) and data have been analyzed according to the procedure described in section 2.9. All liposomes show the same sharp decrease of the scattered intensity $I_{90}$ and, at the same time, a net increase of $R_g$ close to the temperature where the calorimetric transition has been detected. This has to be expected as the solid-to-liquid transition of the membrane triggers a drop of the directional order of the alkyl-chains within the membrane and produces a sudden increase of the single-lipid (intramembrane) diffusion coefficient: as the transition is approached from below, the tilt angles of the chains are increasingly decorrelated in space, producing a drop of the membrane polarizability, an increase of head-group mobility and hence a reduction of its refractive index \cite{michel_determination_2006} as well as the bilayer thickness. Moreover, as the tilt angle of the chains increases on average above $T_c$, being the mass of the bilayer conserved, the size of the liposomes must increase as the transition from solid to melt proceeds. Our data confirms this scenario for all our systems.

The (optical) melting transitions $T_c^{opt}$ have been determined by fitting the time-averaged intensity at $\theta$=$90^\circ$, $I_{90}$(T) with a Boltzmann function \cite{michel_determination_2006}
\begin{equation}\label{Boltzmann}
    I_{90}(T)=(I_s-I_{m})\cdot \left( 1+e^{\frac{T-T_c^{opt}}{\Delta T}} \right){,}
\end{equation}
where $I_s$ is the intensity in the solid phase, $I_m$ is the intensity in the melt phase,
$T_c^{opt}$ the melting temperature, that is half way between the two limiting values $I_s$ and $I_m$, and $\Delta T$ the width of the transition. As described in the manuscript, the same melting temperatures (within 0.2 $^{\circ}$C) and transition widths are obtained if the $R_g(T)$ curves are fitted instead of $I_{90}(T)$ with
\begin{equation}\label{Boltzmann2}
    R_g(T)=(R_s-R_{m})\cdot\left( 1+e^{\frac{T-T_c^{opt}}{\Delta T}} \right){,}
\end{equation}
where $R_s$ and $R_m$ represent the gyration radii in the solid and melted state, respectively.

\begin{table}
\begin{center}
\centering
\footnotesize
\caption{Melting temperatures ($T_c^{opt}$) and gyration radii below and above ($R_s$,$R_m$, respectively) of bare liposomes ($\rho=$0). The transition temperatures are obtained via equation \ref{Boltzmann2}.} \label{tab:table0}
\begin{tabular}{cccc}
$X_{PG}$ & $R_s$ [nm]& $R_m$ [nm] & $T_c^{opt}$\\
\hline
0.33& 49.3 $\pm$ 0.5 & 58.6 $\pm$ 0.5& 46.1 $\pm$ 0.5\\
0.50& 46.4 $\pm$ 0.5 & 56.4 $\pm$ 0.5& 43.9 $\pm$ 1.0\\
0.66& 47.1 $\pm$ 0.7 & 55.3 $\pm$ 0.5& 40.5 $\pm$ 0.9\\
\end{tabular}
\end{center}
\end{table}

As already evidenced by the calorimetric characterization of these systems (see Fig.  3 
 in the manuscript), the bilayer compositions affects the melting transition and a systematic shift to lower temperatures for increasing $X_{PG}$ values is observed (see arrows in Fig. \ref{Irg}). A progressive reduction of the liposome sizes in the solid and in the melt state is also observed (Table \ref{tab:table0}) as the content of DPPG increases. This is expected as HSPC has larger area per molecule than DPPG at the bilayer membrane packing densities \cite{marsh_lateral_1996,dahim_physical_2002}.
Note that since the thermal history of the suspensions probed with LS is not the one used for the DSC (see section 2.8), we do not expect to recover exactly the same melting temperatures. Moreover the two techniques do not physically probe the same transition: LS, in principle, is more sensitive to the order-disorder transition of the polar headgroups, rather than the conformational changes of the hydrophobic chains occurring deep into the membrane, the latter determining most of the excess enthalpy measured by calorimetry. This aspect, interesting \emph{per se}, will not be discussed further.

\begin{figure}[htbp]
  \includegraphics[width=12cm]{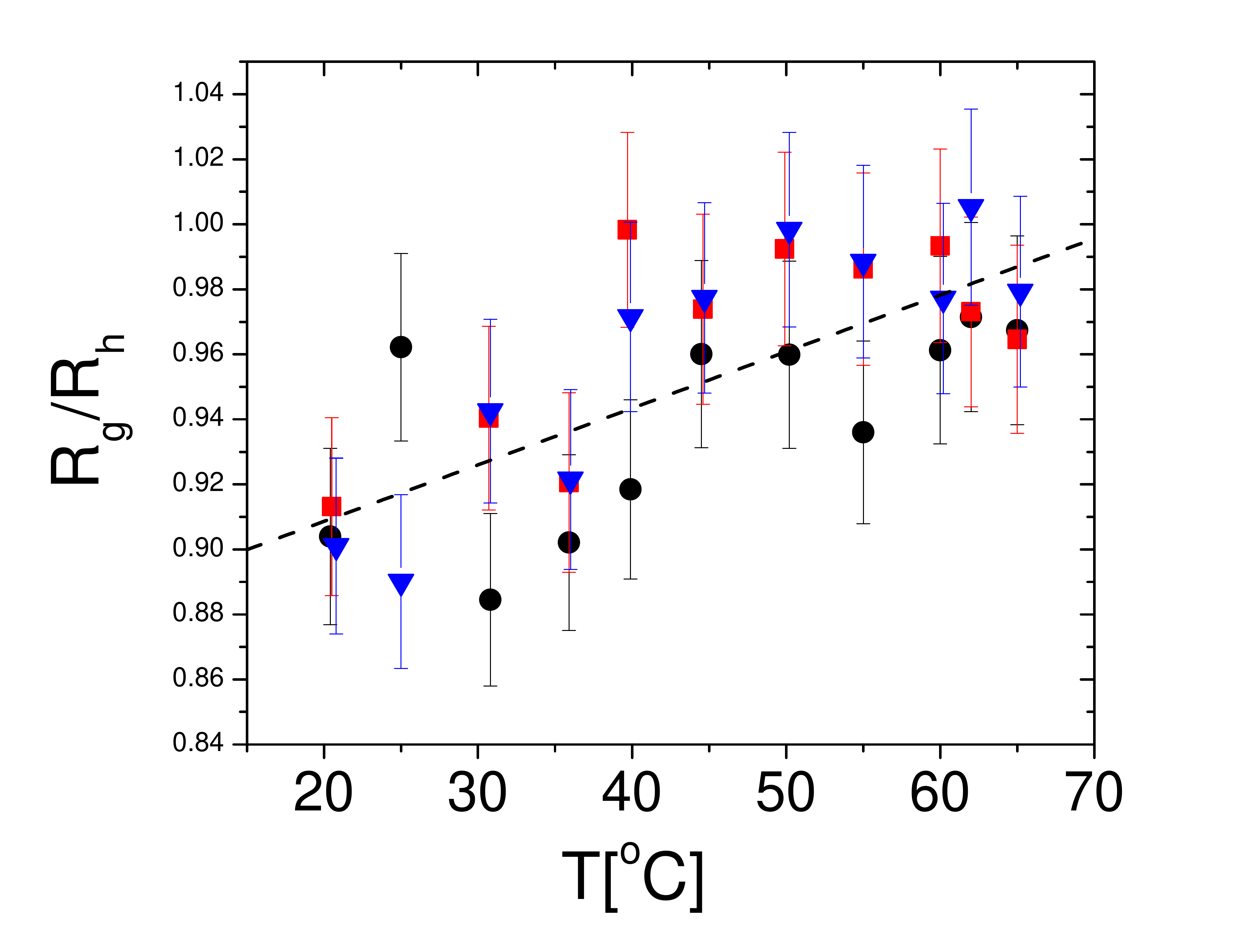}\\
  \caption{Rg/Rh ratio in function of the temperature of the three bare unilamellar mixed liposome preparations $X_{HS}=0.66$ (red squares), $X_{HS}=0.50$ (black circles), $X_{HS}=0.33$ (blue triangles). The dashed black line is the best linear fit obtained including all the data. We obtain $R_g/R_h=(0.00170 \pm 0.00045)T+(0.87 \pm 0.02)$ }\label{rgrh}
\end{figure}

By analysis of the intensity  $I_{90}$ measured as a function of temperature,  we have also extracted the hydrodynamic radius $R_h$ of all unilamellar liposomes via DLS and determined the ratio $R_g/R_h$ which is related to the liposome mass size distribution.
Figure \ref{rgrh} shows the ratio $R_g/R_h$ in function of temperature for all the liposome preparations.  For all temperatures we find $R_g/R_h>0.77$, that is the value expected for homogenous spheres, which points out and confirms that the distribution of mass is peaked at the periphery of the liposomal (hollow) spheres. Despite the data dispersion and the large experimental errors that does not allow discerning any trend difference among the three different mixtures, we do observe an unambiguous increase of the $R_g/R_h$ ratio for increasing temperatures, signaling the shift of liposome mass distributions towards their periphery, and corroborating the scenario where all liposomes increase in size with progressively thinner bilayers.

\section{Monolayer investigations}
Hereafter we report on the surface pressure isotherms of one- and two-component lipid monolayers obtained after the injection in the subphase of different volumes of pure ethanol and INH ethanolic solutions. All the measurements have been performed after an equilibration time that was system-dependent and never less than 600 s.
Figure \ref{Fig:ISO} shows surface pressure isotherms of pure DPPG and HSPC monolayers measured after the injection of pure ethanol and ethanol solutions of INH.
For both lipids, injections of solutions of INH gave rise systematically to a larger pressure increase (larger isotherm shift) with respect to pure ethanol injections, indicating that INH molecules reach the interface, interact with lipid monolayers and actively contribute to decrease surface tension.

\begin{figure}[htbp]
  \includegraphics[width=13cm]{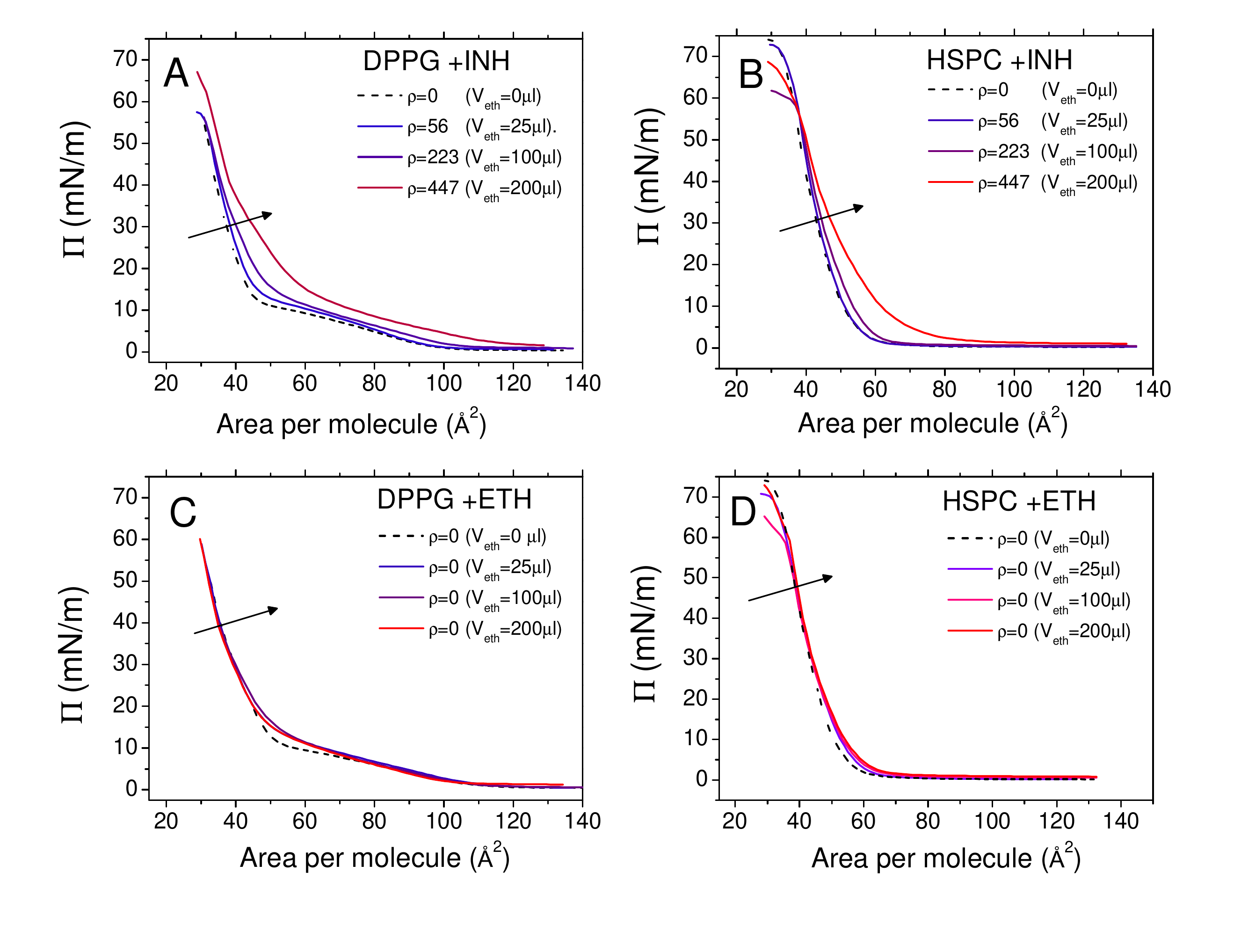}\\
  \caption{Surface pressure isotherms of DPPG (A-C) and HSPC (B-D) monolayers after injection of different volumes of INH or pure ethanol in the subphase. }\label{Fig:ISO}
\end{figure}

For mixed DPPG-HSPC monolayers the scenario is more complex as discussed in section 4.3 of the main manuscript, with the largest differences between injections of INH solutions and pure ethanol at $X_{PG}=$0.33 and $X_{PG}=$0.66, where we also find the minimum and maximum excess free energy of mixed monolayers for 200 $\mu$l of ethanolic INH solution injected in the subphase (Figure 9-F in the main text). The study of mixed monolayers at increasing injected volumes of INH solution, though it is interesting \emph{per se}, goes beyond the scope of our work and we reserve the right to study this issue in the near future.

\begin{figure}[htbp]
  \includegraphics[width=13cm]{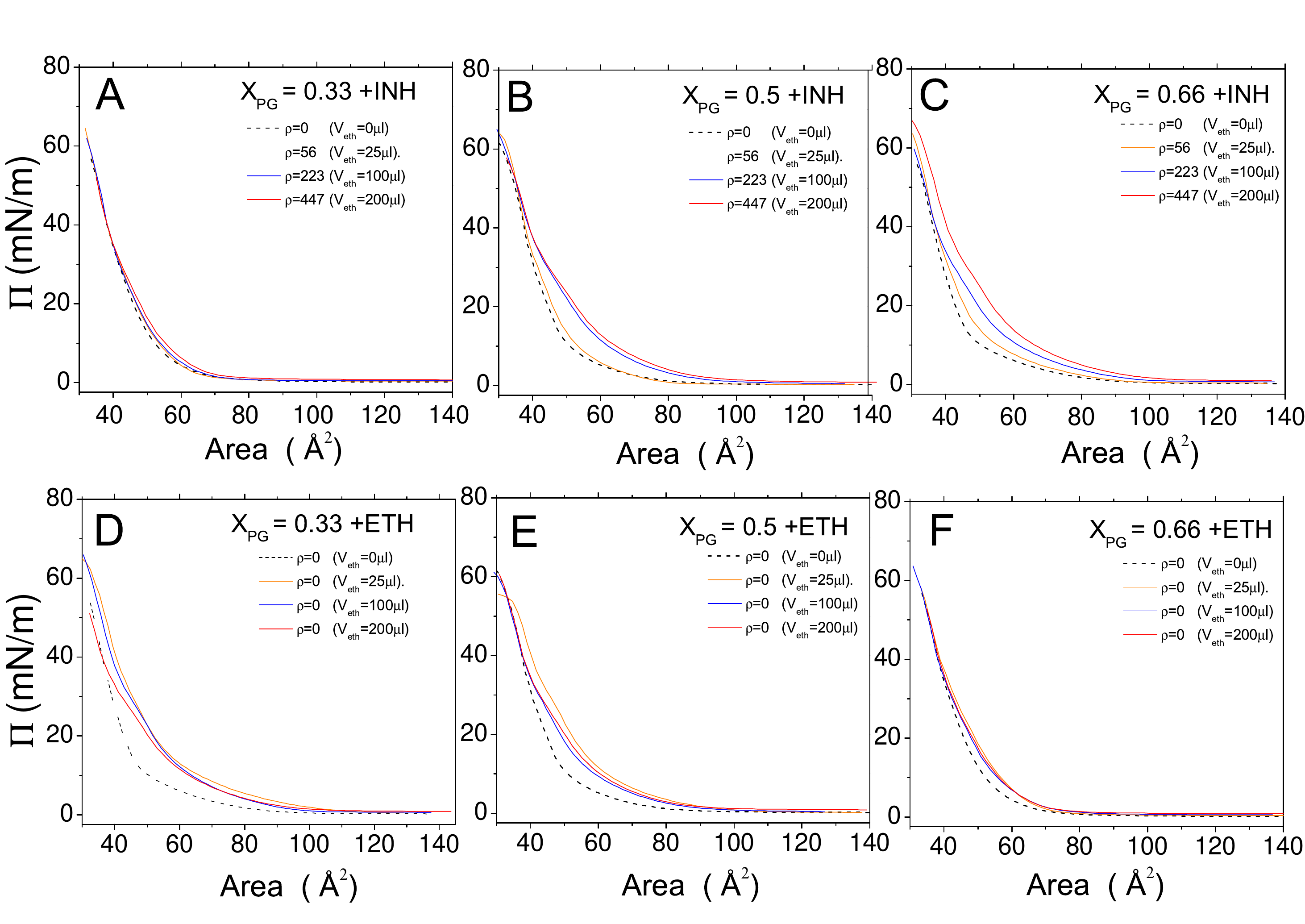}\\
  \caption{Surface pressure isotherms of mixed monolayers with different $X_{PG}$ molar fractions after injection of different volumes of INH or pure ethanol in the subphase.}\label{Fig:ISO}
\end{figure}